\documentclass[aps,pre,reprint,longbibliography,floatfix]{revtex4-2}
\usepackage[utf8]{inputenc}
\usepackage{amssymb,amsmath}
\usepackage{epsfig}
\usepackage{bm}
\usepackage{soul}
\usepackage{xcolor}
\usepackage[mathlines]{lineno}%
\usepackage{epsfig}

\usepackage{graphicx}

\usepackage{siunitx}

\usepackage{dcolumn}
\newcommand{\sby}[1]{}
\newcommand{\sbyt}[1]{{ #1}}
\newcommand{\vicente}[1]{{ #1}}

\newcommand\beq{\begin{equation}}
\newcommand\eeq{\end{equation}}
\newcommand\beqa{\begin{eqnarray}}
\newcommand\eeqa{\end{eqnarray}}

\newcommand{\al}{\alpha}
\newcommand{\rb}{\mathbf{r}}
\newcommand{\vb}{\mathbf{v}}

\newcommand{\la}{\left\langle}
\newcommand{\ra}{\right\rangle}

\begin{document}

\title{Particles, trajectories and diffusion: random walks in cooling granular gases
}

\author{Santos Bravo Yuste}
\email{santos@unex.es}
\homepage{https://fisteor.cms.unex.es/investigadores/\hspace{0pt}santos-bravo-yuste/}
\affiliation{Departamento de F\'{\i}sica and Instituto de Computaci\'on Cient\'{\i}fica Avanzada (ICCAEx), Universidad de Extremadura, E-06006 Badajoz, Spain}
\author{Rub\'en G\'omez Gonz\'alez}
\email{ruben@unex.es}
\affiliation{Departamento de Did\'actica de las Ciencias Experimentales y las Matem\'aticas, Universidad de Extremadura, E-10004 C\'aceres, Spain}
\author{Vicente Garz\'o}
\email{vicenteg@unex.es}
\homepage{\hspace{0pt}https://fisteor.cms.unex.es/investigadores/\hspace{0pt}vicente-garzo-puertos/
} 
\affiliation{ Departamento de F\'{\i}sica and Instituto de Computaci\'on Cient\'{\i}fica Avanzada (ICCAEx), Universidad de Extremadura, E-06006 Badajoz, Spain}

\begin{abstract}	
We study the mean-square displacement (MSD) of a tracer particle diffusing in a granular gas  of inelastic hard spheres under homogeneous cooling state (HCS). Tracer and granular gas particles are in general  mechanically different. Our approach uses a series representation of the MSD where the $k$-th term is given  in terms of the mean scalar product $\langle \rb_1\cdot\rb_k \rangle$,   with $\rb_i$ denoting the displacements of the tracer between successive collisions.  We find that this series approximates a geometric series with the ratio $\Omega$. We derive an explicit analytical expression of $\Omega$ for granular gases in three dimensions, and validate it through a comparison with the numerical results obtained from the direct simulation Monte Carlo (DSMC) method. Our comparison covers a wide range of masses, sizes, and  inelasticities. 
From the geometric series, we find that the MSD per collision is simply given by the mean-square free path of the particle divided by $1-\Omega$.  The analytical expression for the MSD derived here is compared with DSMC data and with the first- and second-Sonine approximations  to the MSD obtained from the Chapman-Enskog solution of the Boltzmann equation. Surprisingly, despite their simplicity, our results outperform  the predictions of the first-Sonine approximation to the MSD, achieving accuracy comparable to the second-Sonine approximation.
\end{abstract}

\date{\today}
\maketitle


\section{Introduction}
\label{sec1}

In two articles published in 1906 (one on the kinetic theory of diffusion in molecular gases \cite{[][{. English translation in: Marian Smoluchowski, Selected Scientific Works, editor Bogdan Cichocki,  Wydawnictwa Uniwersytetu Warszawskiego, Warsaw 2017, pp. 77-86.}]Smoluchowski1906a} and the other on the kinetic theory of Brownian motion \cite{[{}][{. English translation in: Marian Smoluchowski, Selected Scientific Works, editor Bogdan Cichocki,  Wydawnictwa Uniwersytetu Warszawskiego, Warsaw 2017, pp. 87-106.}]Smoluchowski1906b}; the former complementing Einstein's famous 1905 article on the same subject \cite{Einstein1905}), Smoluchowski pioneered the use of tools from what is now recognized as random walk theory (or stochastic processes) to determine, among other things, the diffusion coefficient of a tracer (or intruder) particle in a molecular gas. Smoluchowski's results are based on two  main  assumptions. First, it is assumed that if the tracer particle is immersed in a dilute gas, its mean free path is much larger than the size of the gas molecules. Second, it is further assumed that the tracer particle scatters isotropically after colliding with other molecules (an assumption that is not acceptable even to a first approximation if the tracer is Brownian, namely, when it is much more massive than the gas molecules). 
Under these assumptions, the MSD of the tracer particle, $\la R^2\ra$, after $N$ collisions is given by 
\beq
\label{0.1}
\la R^2\ra =2N \la r\ra^2,
\eeq
where  $\la r\ra$ is the mean free path ($r$ is the free path). Since $\la r\ra=\la v \ra \tau$ (where $\tau=t/N$ is the mean free time between collisions during time $t$  and  $\la v\ra $ is the mean velocity, with $v$ being the velocity of the particle between two successive collisions), then it follows that 
\beq
\label{0.2}
\la R^2\ra=2\la r\ra \la v\ra t.
\eeq
Equation \eqref{0.2} leads to the diffusion coefficient
\begin{equation}
	\label{DeKT}
D=\frac{\la R^2\ra}{6t}=	\frac{\langle r \rangle \langle v \rangle}{3} \equiv D_\text{eKT}.
\end{equation}

Expression \eqref{DeKT} is nothing more than the expression of the diffusion coefficient given by the elementary kinetic theory, traceable back to Boltzmann, and found in standard textbooks \cite{Reif1965}. 
In Ref.~\cite{Smoluchowski1906b}, Smoluchowski further noted that the tracer particle, after colliding with a gas molecule (especially if the tracer particle is heavy), tends to be scattered in a direction similar to the direction it was carrying before the collision. This is known as persistence of the collisions. This fact explains why the elementary formula \eqref{DeKT} is only a first approximation. Smoluchowski dared in Ref.\  \cite{Smoluchowski1906b} to make an estimate of the correction to the MSD due to persistence by introducing in his analysis the mean value of the angle at which the direction of motion of the particle changes in each collision. As Smoluchowski himself pointed out (section 7 of Ref.~\cite{Smoluchowski1906a}), Jeans had already observed in 1904 \cite{Jeans1904} that consideration of the existence of persistence in collisions was necessary for a better microscopic explanation of the properties of gases and, in particular, for a better estimate of the tracer diffusion coefficient \cite{Jeans1940}.

To the best of our knowledge, the first rigorous analyisis of the MSD of a particle in a gas by means of a random walk approach was addressed by Yang in 1949 \cite{Yang1949}. The MSD is defined as 
\beq
\label{0.3}
\la R^2\ra=\la\mathbf{R}\cdot\mathbf{R}\ra=\sum_{i=1}^{N}\sum_{j=1}^N \langle \rb_i\cdot\rb_j \rangle,
\eeq
where $\rb_j$ are the displacements of the particle between successive collisions and $\mathbf{R}=\sum_{i=1}^{N} \rb_i$ is the displacement of the particle after $N$ collisions. 
  (As we mention below, although Yang's estimates for  the first few terms $\la\rb_i\cdot\rb_j\ra$  for elastic hard spheres were quite inaccurate, he did provide the expressions needed to determine them.) 
In particular, if the persistence in collisions is neglected, then $\la\rb_i\cdot\rb_j\ra=0$ when $i\neq j$ and the MSD 
reduces to the elementary expression $\la R^2\ra=N \la r^2\ra$, where $\la r^2\ra\equiv\la r_i^2\ra$ denotes the mean value of the square of the free path (MSD between two successive collisions). In terms of the mean free time $ \tau=t/N$, the MSD can be rewritten as $\la R^2\ra= \la r^2\ra t/ \tau$, which yields the following expression of the diffusion coefficient $D$:
\begin{equation}
	\label{DeRW}
	D=\frac{\la R^2\ra}{6t}=\frac{ \la r^2\ra}{6\tau}\equiv D_\text{eRW}.
\end{equation}

While very similar, this expression $D_\text{eRW}$ for the diffusion coefficient  is not exactly equal  to $D_\text{eKT}$.  This is because $\la r \ra \la v \ra$ is not exactly equal to $\la r^2\ra/(2\tau)=\la r v \ra$ (see Sec.\ V of Ref.\ \cite{Yuste2024}).
Just as Smoluchowski in Ref.\ \cite{Smoluchowski1906b} improved the estimation of the diffusion coefficient by including the effects of persistence, the elementary expression \eqref{DeRW} for $D$ can also be improved if the persistence in collisions (which implies $\la\rb_i\cdot\rb_j\ra\neq 0$) is accounted for. Unfortunately, as said before, Yang's estimate of the terms with $i\neq j$ for a gas of elastic hard spheres was deficient. This is understandable given the computational limitations in the late 1940s when Yang carried out the calculations for his article. The first three  terms with $i\neq j$ (i.e., $\la\rb_1\cdot\rb_2\ra$, $\la\rb_1\cdot\rb_3\ra$ and $\la\rb_1\cdot\rb_4\ra$)  for the case of an elastic hard sphere gas have recently been calculated in Ref.\ \cite{Yuste2024}. In this paper, it was observed that the corrections due to the persistence of collisions to the elementary value $D_\text{eRW}$ of the diffusion coefficient decay, in a very good approximation, exponentially. The factor by which these terms decrease is very well approximated by the quantity known in kinetic theory of gases as the mean-persistence ratio $\la \omega \ra$ \cite{CCC70}. This result led us to propose in Ref.\ \cite{Yuste2024}  the following expression for the MSD after $N$ collisions:
\begin{equation}
	\label{R2Nelastico}
	\la R^2\ra=N \frac{\la r^2\ra}{1-\la \omega \ra}.
\end{equation}
From Eqs.\ \eqref{DeRW} and \eqref{R2Nelastico}, and $N=t/\tau$, one  gets 
\begin{equation}
	\label{DRWelastico}
	D=\frac{D_\text{eRW}}{1-\la \omega \ra}.
\end{equation}
The expression \eqref{R2Nelastico} for the diffusion coefficient (or equivalently for the MSD) substantially improve the corresponding elementary expressions. In fact, they provide results comparable to the standard results of the kinetic theory of gases derived from the Chapman-Enskog solution of the Boltzmann equation \cite{Yuste2024} .

The aim of the present work is to generalize the arguments used in Ref.~\cite{Yuste2024} for calculating the MSD in molecular gases to \emph{granular} gases. It is well known that, under rapid flow conditions, a granular gas can be modeled as a gas of hard spheres with \emph{inelastic} collisions. If the spheres are assumed to be perfectly smooth, the inelasticity in collisions is accounted for only by a constant coefficient of normal restitution $0<\alpha\leq 1$. When $\alpha=1$, the collisions are elastic. In a binary collision in the inelastic hard sphere model,  the tangential component of the relative velocity of the two colliding spheres remains unchanged, while its normal component is shrunk by a factor $\alpha$. Since in our problem the tracer and the particles of the granular gas are in general mechanically different, there are two different coefficients of restitution: $\alpha$ (associated with the collisions between the granular particles themselves) and $\alpha_0$ (associated with the collisions between the tracer and the particles of the granular gas).  
As in Ref.\  \cite{Yuste2024}, our goal here is to determine the diffusion of a tracer particle moving in a freely cooling granular gas.  In this \emph{homogeneous} state (which is referred to as the HCS in the granular literature \cite{BP04,GarzoBook19}) the grains collide freely and cool accordingly. However, one of the most characteristic features of granular gases (as compared to conventional molecular gases) is the spontaneous formation of velocity vortices and density clusters in freely cooling flows \cite{GZ93,M93}. The instability of the HCS can be well described by a linear stability analysis of the granular version of the Navier--Stokes hydrodynamic equations \cite{BDKS98,BRC99,G05,GMD06,MGHEH12,BR13,MGH14,G15}. The analysis clearly shows that the origin of the above instability is related to inelasticity in collisions, and since it is limited to small wave numbers, it can be avoided for sufficiently small systems. The study of diffusion in freely cooling granular gases provided in this paper is limited to situations where the HCS is stable and thus the granular gas maintains its homogeneous state.

As previously mentioned, in this article we analyze the diffusion of a tracer granular particle in a granular gas under HCS using a procedure that views the motion of the tracer particle as a random walk. It is important to remark that this procedure for inelastic collisions is fundamentally different from the one followed in granular kinetic theory \cite{GarzoBook19} when one extends the conventional Chapman--Enskog method \cite{CCC70} to granular gases. In this paper,
we will compare our findings derived from random walk arguments with (i) previous theoretical results \cite{GM04,GV09}  
for the tracer diffusion coefficient derived from the Chapman--Enskog solution in the so-called first and second Sonine approximations, and with (ii) our computer simulation results obtained from the DSMC method \cite{Bird94}.

\sby{Respuesta a Reviewer 3, punto 1:}
\vicente{
	It is important to note that, although  this paper focuses on validating our random walk method primarily via an analysis of the MSD for long times, our procedure also  describes (see Sec.~\ref{sec:MSDregimes}) the MSD behavior across all temporal regimes: ballistic at very short times, diffusive  or subdiffusive at intermediate times and subdiffusive at long times (provided inelastic collisions exist).
}

The organization of the article is as follows. 
In Sec.\ \ref{sec:HCS}, we describe the system (granular gas plus tracer particle) we are interested in and give some known relationships that will be used throughout this paper.
The MSD of the tracer particle in terms of its elementary displacements (or free paths) $\rb_i$ is given in Sec.~\ref{sec:RW} and evaluated in Sec.\ \ref{sec:SmoJeans}.   In this section  it is shown that the MSD can be approximated by a geometric series (which we call the collisional series) where the ratio between the successive terms is expressed by means of averages that involve the velocities before and after the collision and the scattering angles. For inelastic collisions, this ratio is the mean-persistence ratio $\la \omega \ra$ of kinetic theory of gases.
In Sec.~\ref{sec:Omega}, we provide a theoretical expression, denoted by $\Omega$, to estimate the collisional series ratio for a tracer particle moving in a granular gas in HCS.
Computer simulation results obtained from the DSMC method \cite{Bird94} are presented in Sec.\ \ref{sec:MSDsimu}. These simulation data are used 
to validate that the collisional series closely approximates a geometric series and that $\Omega$ accurately predicts its ratio. 
We also compare the MSD derived from $\Omega$ with simulation data, finding surprisingly good agreement between theory and simulation.
\sbyt{
We show that this is due to the cancellation of errors in the auxiliary (or intermediate) approximations employed to arrive at the final expressions for $\Omega$ and the MSD.
}
In Sec.~\ref{sec:Boltzmann}, we further compare  the theoretical and computational results derived here with those previously obtained \cite{GM04} by solving the Boltzmann--Lorentz equation by means of the Chapman--Enskog method \cite{CCC70}. 
We end the paper in Sec.~\ref{sec:Conclu} where we collect some conclusions, remarks, and possible extensions of this work.

\section{Tracer particle moving in a granular gas under HCS}
\label{sec:HCS}

We consider a physical system consisting of a tracer (or intruder) particle immersed in a granular gas. Both the tracer and particles of the granular gas are modeled as a gas of hard spheres with inelastic collisions. The intruder may be identical to or different from the gas particles (grains). The granular gas is assumed to be in HCS, namely, a homogeneous state where the granular temperature $T$ decreases in time. Detailed descriptions and relevant equations are provided in Refs.\ \cite{GarzoBook19,Abad2022}.

We denote the mass and diameter of the granular gas particles as $m$ and $\sigma$, respectively, and those of the intruder as $m_0$ and $\sigma_0$. Throughout this article, we use the convention that intruder-related quantities are indicated by a subscript 0.   Accordingly,  the inelasticity of collisions between  granular gas particles is quantified by the coefficient of normal restitution $\alpha$, while the inelasticity for intruder-gas collisions is characterized by the coefficient of normal restitution $\alpha_0$.

Due to the inelastic character of collisions, the mean kinetic energy (granular temperature) of the granular gas decays with  time. Its evolution equation is \cite{BP04,GarzoBook19}
\beq
\label{2.1}
\partial_t \ln T(t)=-\zeta(t),
\eeq
where $\zeta$ is the cooling rate. This quantity gives the rate of energy dissipation due to inelastic collisions. Since for inelastic hard spheres  $\zeta(t)\propto \sqrt{T(t)}$, the integration of Eq.\ \eqref{2.1} yields Haff's law \cite{H83}:
\begin{equation}
	\label{HaffLaw}
	 T(t)=\frac{T(0)}{\left(1+t/t_\zeta\right)^2}, 
\end{equation}
where $T(0)$  is the initial temperature, $t_\zeta=2/\zeta(0)$ is the characteristic cooling time,  and $\zeta(0)$ is the cooling rate  at $t=0$. The partial temperature $T_0(t)$ of the tracer particle is a relevant quantity at a kinetic level as it provides a measure of its mean kinetic energy. In the HCS, $T_0(t)\propto T(t)$ but the temperature ratio $T_0(t)/T(t)$ remains constant in time and, in general, is different from 1 (breakdown of kinetic energy \vicente{equipartition} assumption) \cite{GD99b}. The condition for determining the temperature ratio is $\zeta(t)=\zeta_0(t)$, where $\zeta_0$ is the partial cooling rate associated with $T_0$. Given that both cooling rates are given in terms of the distribution functions of the granular gas and the tracer particles (whose exactly forms are not known to date), one usually approximates these distributions by their Maxwellian forms. In this approximation and in terms of the auxiliary parameter 
\beq
\label{2.2}
\beta=\frac{m_0 T}{m T_0},
\eeq
the temperature ratio is the unique positive root of the equation \cite{GD99b,Abad2022}:
\begin{align}
	\frac{1-\alpha^2}{d}=&\frac{2\sqrt{2}}{d} \mu \frac{\chi_0}{\chi} \left(\frac{\bar{\sigma}}{\sigma}\right)^{d-1} \left(\frac{1+\beta}{\beta}\right)^{1/2}(1+\al_{0})\nonumber\\
	&\times
	\left[1-\frac{1}{2}\mu (1+\beta)(1+\al_{0})\right].
\label{ecubeta}
\end{align}
Here, $d$ is the dimensionality of the system ($d=3$ for spheres and $d=2$ for disks), and $\chi$ and $\chi_0$ are the pair correlation functions at contact for the granular gas and the intruder-gas particles, respectively. Moreover,  
\begin{align}
\label{mu}
	\mu&=\frac{m}{m+m_0},\\
\label{sigma}
	\bar{\sigma}&=\frac{\sigma+\sigma_0}{2}.
\end{align}
Note that Eq.~\eqref{ecubeta} can be converted into a cubic equation for $\beta$, allowing for an explicit, albeit cumbersome, solution. Nevertheless, the numerical resolution of Eq.~\eqref{ecubeta} is straightforward and efficient.
%
\sby{Respuesta a Reviewer 1 punto 1 y Reviewer 3 punto 2:}
\vicente{The parameter $\beta$ derived from  the physical solution to  Eq.~\eqref{ecubeta} is time-independent because all coefficients in that equation are constant. This means the ratio  $T(t)/T_0(t)$ given by Eq.~\eqref{2.2} is also time-independent  in the HCS. Consequently,  $T(t)$ y $T_0(t)$  share the same temporal dependence, which for $T(t)$ is  given by Eq.~\eqref{HaffLaw} while for $T_0(t)$ is}  
\begin{equation}
\label{HaffLaw0}
		T_0(t)=\frac{T_0(0)}{\left(1+  t/t_\zeta\right)^2}.
\end{equation}

A decrease in the gas temperature (mean kinetic energy and, consequently, mean velocity) implies that the (average) collision frequencies $\nu(t)$ and $\nu_0(t)$ for grain-grain and intruder-grain collisions, respectively, are decreasing functions of time. The collision frequency $\nu(t)$ is given by  \cite{GarzoBook19} 
\begin{equation}
\label{nut}
\nu(t)=\frac{\sqrt{2}\pi^{(d-1)/2}}{\Gamma\left(\frac{d}{2}\right)} n \sigma^{d-1}\chi v_\text{th}(t)
\end{equation}
where $n$ denotes the number density of the granular gas particles and 
\begin{equation}
	\label{vth}
	v_\text{th}=\sqrt{\frac{ 2T(t)}{m}}
\end{equation}
is the thermal velocity. The expression \eqref{nut} for $\nu$  
is identical to that for molecular (elastic) gases, except that for granular gases the thermal velocity decreases over time due to the inelastic nature of collisions.
The intruder-grain collision frequency $\nu_0(t)$ is well approximated by the relation \cite{Abad2022}
\begin{equation}
\label{nu0nu}
\nu_0(t)=\Upsilon \,\nu(t)
\end{equation}
where 
\begin{equation}
\label{Upsilon-def}
\Upsilon=\left(\frac{\bar{\sigma}}{\sigma}\right)^{d-1}\frac{\chi_0}{\chi}\left(\frac{1+\beta}{2\beta}\right)^{1/2}.
\end{equation}

From the collision frequency $\nu_0(t)$, we obtain the average number of collisions in time $t$: 
\beq
\label{2.3}
s_0(t)=\int_0^t \nu_0(t')dt'=\Upsilon s(t),
\eeq
where 
\beq
\label{2.4}
s(t)=\int_0^t \nu(t') dt'
\eeq
is the mean number of collisions experienced by a granular gas particle with the other particles of the granular gas during time $t$. An explicit expression for $\nu(t)$ is obtained by using Haff's law \eqref{HaffLaw} in the thermal velocity expression \eqref{vth}. Integration of Eq.\ \eqref{nut}  yields
\begin{equation}
	\label{sself}
	s(t)=\frac{2\nu(0)}{\zeta(0)} \ln\left(1+\frac{\zeta(0)}{2\nu(0)}\, t^*\right)	 
		=\frac{t_\zeta}{\tau(0)}  \ln\left(1+  t/t_\zeta\right)
\end{equation}
where  $t^*=\nu(0) t=t/\tau(0)$  is the time in units of the initial intercollisional time   $\tau(0)=\nu(t=0)^{-1}$  of the gas particles. 
From Eqs.~\eqref{2.3} and \eqref{sself} we obtain the mean number of intruder collisions up to time $t$:
	\begin{equation}
		\label{s0t}
		s_0(t)=\frac{t_\zeta}{\tau_0(0)}  \ln\left(1+  t/t_\zeta\right)
	\end{equation}	
	where  $\tau_0(0)=1/\nu_0(0)$ is the initial mean collision time between the intruder and the gas particles. A good approximation for   $t_\zeta$  is given by   $t_\zeta/\tau(0)=2d/(1-\alpha^2)$ \cite{Abad2022}, which, from Eq.~\eqref{nu0nu}, implies 
	\begin{equation}
		\label{tzetatau0}
		\frac{t_\zeta}{\tau_0(0)}= \Upsilon \frac{t_\zeta}{\tau(0)}=\frac{2d \Upsilon}{1-\alpha^2}.
	\end{equation}

On the other hand, the Maxwell mean free path \cite{Paik2014} $\ell_\text{M}$ is defined as
\beq
\label{2.5}
\ell_\text{M}\equiv\frac{\bar{v}(t)}{\nu(t)},
\eeq
where $\bar{v}\equiv \la v\ra$ denotes the average speed modulus. The Maxwell mean free path of the particles of the granular gas has the same form as for conventional molecular (elastic) gases: 
\begin{equation}
	\label{eleM}
	\ell_\text{M}=\frac{\bar{v}(t)}{\nu (t)}=\frac{\Gamma\left(\frac{d+1}{2}\right)}{\sqrt{2}\pi^{(d-1)/2}}\frac{1}{n\sigma^{d-1}\chi}.
\end{equation}
Analogoulsy, the Maxwell mean free path of the intruder, $\ell_{0\text{M}}\equiv\bar{v}_0(t)/\nu_0(t)$, is given by  \cite{Abad2022}
\begin{equation}
	\label{ele0M}
	\ell_{0\text{M}}=\frac{\ell_{\text{M}}}{\Upsilon \sqrt{\beta}}.
\end{equation}

\section{The intruder as a random walker. Mean square displacement}
\label{sec:RW}

Taking the intruder's position $\mathbf{R}(t=0)$ immediately after a collision as the initial position and this collision time as $t=0$, the intruder's position after $N$ collisions is $\mathbf{R}_N= \sum_{i=1}^N \rb_i$, where $\rb_i$ is the $i$-th displacement between the $i$-th and $(i-1)$-th collisions (with the  initial collision labeled as 0). As usual, the intruder diffusion is quantified by the  variance of its position after time $t$. With no external forces, the mean position is $\langle \mathbf{R}(t)\rangle=0$, so the variance is simply $\langle R^2\rangle$.

To evaluate the intruder's position variance $\langle R^2(t)\rangle$ after time $t$, we begin by considering the variance after $N$ displacements (or equivalently, steps or collisions):
\begin{align}
\label{MSD2}
 \la {R}^2_N\ra
  =&
  \langle \mathbf{R}_N\cdot \mathbf{R}_N\textbf{}  \rangle   \nonumber\\
  =&
   \sum_{i=1}^N  \la \rb_i  \cdot \rb_i \ra
  +  \, 2 \sum_{i=1}^{N-1}   \la \rb_i  \cdot \rb_{i+1} \ra
  +  \, 2 \sum_{i=1}^{N-2}   \la \rb_i  \cdot \rb_{i+2} \ra 
 \nonumber \\
  &+\cdots
\end{align}
In a HCS state, free paths between collisions are time-independent (see Sec.~\ref{sec:Boltzmann}), so that  
\begin{eqnarray}
	\label{MSDles2}
	\langle {R}^2_N\rangle
	&=&
N\langle \rb_1\cdot \rb_{1} \rangle+2(N-1) \langle \rb_1\cdot \rb_{2} \rangle+2(N-2)\langle \rb_1\cdot \rb_{3} \rangle\nonumber\\
& & +\cdots \nonumber\\
&=& 
	N \langle \rb_1\cdot \rb_{1} \rangle  +
	2\sum_{k=1}^{N-1} (N-k) \langle \rb_1\cdot \rb_{1+k}\rangle.
\end{eqnarray}
Equation \eqref{MSDles2} can be rewritten as
\begin{equation}
	\label{MSDle3}
	\langle {R}^2_N\rangle=N\ell_e^2
\end{equation}
where $\ell_e$ is the effective free path. This name is justified because the MSD of a walk with $N$ isotropic, uncorrelated steps of fixed size $\ell_e$ is exactly given by Eq.~\eqref{MSDle3}.
Note that $\ell_e^2$ is the MSD of the intruder per collision. Comparing Eq.~\eqref{MSDle3} with Eq.~\eqref{MSDles2}, the effective free path can be approximated by
\begin{equation}
	\label{ell2Sume}
	\ell_e^2\approx \langle r_1^2 \rangle + 2\langle \rb_1\cdot \rb_{2}\rangle + 2\langle \rb_1\cdot \rb_{3}\rangle +\cdots.
\end{equation}
for large $N$. 
Upon writing Eq.\ \eqref{ell2Sume} we have assumed that steps  $\rb_{1}$ and $\rb_{k}$ decorrelate rapidly enough that the error in the approximation
\begin{equation}
	\label{apdefast}
	(N-k-1) \,\langle \rb_1\cdot \rb_{k}\rangle \approx N \langle \rb_1\cdot \rb_{k}\rangle
\end{equation}
is negligible even for large $k$. This assumption holds except for extreme cases where the mass ratio $m_0/m$ is large and the diameter ratio $\sigma_0/\sigma$ is small (see Sec.~\ref{sec:MSDsimu}).
In fact, we will show that, to a very good approximation, these correlations decay exponentially, with $\langle \rb_1\cdot \rb_{k}\rangle$ tending towards zero as $k$ increases. This point will be discussed in Sec.~\ref{sec:MSDsimu}.

It is convenient to rewrite Eq.~\eqref{ell2Sume} as
\begin{eqnarray}
\label{ell2redSer}
\frac{\ell_e^2}{\langle  r^2  \rangle}=\frac{\la R^2_N \ra}{N \langle  r^2  \rangle}&\approx&   1+ \frac{2\langle \rb_1\cdot \rb_{2}\rangle}{\langle  r^2  \rangle}+   \frac{2\langle \rb_1\cdot \rb_{3}\rangle}{\langle  r^2  \rangle}+\cdots \nonumber\\
&\equiv& \sum_{k=1}^\infty c_k,
\end{eqnarray}
where we have defined
\begin{subequations}
	\label{ckDef}
	\begin{align}
		c_1 &= 1,\label{ceq1}\\
	c_k&\equiv 2 \frac{\langle \rb_1\cdot \rb_{k}\rangle}{\la r^2\ra},  \quad k\ge 2. \label{ckge2}
	\end{align}
\end{subequations}
The notation $\la r^2\ra$ is introduced as a simplification of $\la r_1^2\ra$, i.e.,  $\la r^2\ra\equiv\la r_1^2\ra$. 
We call  collisional series to the series in Eq.~\eqref{ell2Sume} and  \emph{reduced} collisional series to  $\sum_{n=1}^\infty c_n$.  In summary, Eqs.~\eqref{MSDle3} and \eqref{ell2redSer} tell us that the MSD of the intruder per step, $\langle {R}^2_N\rangle/N$, is the mean square free path, $\langle  r^2  \rangle$, corrected by the reduced collisional series.

As stated in Sec.\ \ref{sec1}, neglecting collision persistence (i.e., assuming isotropic post-collision scattering) leads to $\la\rb_i\cdot\rb_j\ra=0$ for $i\neq j$. In this approximation, one gets  the elementary MSD $\la R^2_N\ra=N \la r^2\ra$. Thus, the reduced collisional series corrects the elementary MSD to yield the correct MSD. In this context, the reduced collisional series quantify the impact of the collision persistence on the MSD of the intruder.

Finally, to relate $\langle R^2(t) \rangle$ and $\langle R^2_N \rangle$, we consider the probability density function (pdf) of the intruder's position $\mathbf{R}$ at time $t$:
\begin{equation}
\label{PR}
	P(\mathbf{R},t)=\sum_{N=0}^\infty P_N(\mathbf{R}) \xi_N(t).
\end{equation}
Here, $P_N(\mathbf{R})$ is the pdf of the position of the walker after $N$ steps and $\xi_N(t)$ is the probability of taking exactly $N$ steps by time $t$. Substituting Eq.~\eqref{PR} into the relation 
\beq
\label{3.1}
\langle R^2(t) \rangle =\int P(\vicente{\mathbf{R}},t) R^2\, d\mathbf{R}
\eeq
and swapping the summation and integration order, we get
\begin{align}
\label{RTRN}
	\langle R^2(t) \rangle=\sum_{N=0}^\infty \xi_N(t) \int P_N(\mathbf{R}) R^2\, d\mathbf{R}
	=\sum_{N=0}^\infty \xi_N(t) \langle R^2_N \rangle.
\end{align}
Using Eq.~\eqref{MSDle3}, we arrive at
\begin{equation}
	\label{MSDelle}
	\langle R^2(t) \rangle
	=\ell_e^2\sum_{N=0}^\infty \xi_N(t) N =\langle N(t)\rangle \,\ell_e^2 \equiv s_0(t) \,\ell_e^2,
\end{equation}
where $\langle N(t)\rangle$ is the average number of steps (collisions) taken by the random walker (the intruder) up to time $t$. We denoted $\langle N(t)\rangle$ as $s_0(t)$ in Sec.~\ref{sec:HCS}.

\section{MSD as a geometric series}
\label{sec:SmoJeans}

Since $s_0(t)$ is known (see Eq.~\eqref{s0t}), Eq.~\eqref{ell2Sume} indicates that to evaluate $\langle R^2(t) \rangle$ we need only  to evaluate $\ell_e^2$, i.e.,  the averages  $\langle \rb_1\cdot \rb_{k}\rangle$. Exact evaluation of these quantities, even for small $k$ and  molecular (elastic) gases, is challenging (see Ref.~\cite{Yuste2024} for more details).
Unfortunately, the method used in Ref.~\cite{Yuste2024} to rigorously compute these quantities cannot be used for  granular gases because detailed balance is not satisfied when the collisions between the particles are inelastic. Nevertheless, Ref.~\cite{Yuste2024} showed that the method followed in this paper built on reasonable approximations produced an estimate of $\langle \rb_1\cdot \rb_{k}\rangle/\la r^2\ra $ that was in excellent agreement with the exact values. These approximations implied that the reduced collisional series \eqref{ell2redSer} formed a geometric series (or, in other words, that $\langle \rb_1\cdot \rb_{k}\rangle$ decays exponentially), which led to the simple (and very accurate) expression given in Eq.~\eqref{R2Nelastico}.  Here, we will extend this approximation to granular gases, recovering the results of Ref.~\cite{Yuste2024} in the elastic limit.

Therefore, we need to evaluate the terms $2\langle \rb_1\cdot \rb_{k}\rangle/\la r^2\ra$ for $k\ge 1$. To do this, we will follow similar steps to those used in Ref.~\cite{Yuste2024} when collisions are elastic.
We start by assuming that the lengths of the first step, $r_1$, and the $k$-th step, $r_k$, along with the angle $\theta_{1,k}$ between $\rb_1$ and $\rb_k$, are mostly uncorrelated. This leads to
\begin{align}
\frac{2\la \rb_1\cdot \rb_{k}\ra}{\la r^2\ra}
&=\frac{2 \la r_1  r_k  \cos \theta_{1,k} \ra}{ \la r^2 \ra} \nonumber\\
&\approx  \frac{2 \la r_1 \ra \la r_k \ra \la \cos \theta_{1,k} \ra}{ \la r^2 \ra}
= \frac{2 \la r \ra^2 \la \cos \theta_{1,k} \ra}{ \la r^2 \ra}.
  \label{r1rkA}
\end{align}
Note that $\la r_1\ra\equiv \la r\ra$ is the mean free path of the intruder, which we will also denote by $\ell_0$. A simple approximation for $\la r^2 \ra$ is $\la r^2 \ra\approx 2\ell_0^2$ (see Sec.~\ref{sec:Boltzmann}). Thus, Eq.~\eqref{r1rkA} becomes
\begin{equation}
	\label{r1rkAb}
	\frac{2\la \rb_1\cdot \rb_{k}\ra}{\la r^2\ra}
	\approx  \la \cos \theta_{1,k} \ra.
\end{equation}
The angles $\theta_{1,k}$,  $\theta_{1,2}$ and $\theta_{\vicente{2},k}$ are related by the spherical cosine law:
\begin{equation}
\label{cos1k}
\cos\theta_{1,k}=\cos\theta_{1,2}\,\cos\theta_{k-1,k} + \sin\theta_{1,2}\sin\theta_{k-1,k} \cos\varphi_{1,2,k},
\end{equation}
where  $\varphi_{1,2,k}$ is the angle between the plane  generated by the vectors $\rb_1$ and $\rb_2$,  and the plane generated by the vectors $\rb_{\vicente{2}}$  and  $\rb_{k}$.
Due to the rotational symmetry of collisions along the direction of the precollisional displacement $\rb_1$ (or velocity $\vb_1$), all values of  $\varphi_{1,2,k}$ are equally probable.  Therefore 
$ \langle   \sin\theta_{1,2}\sin\theta_{\vicente{2},k} \cos\varphi_{1,2,k} \rangle=0$, and
 \begin{equation}
\la    \cos\theta_{1k}\ra=\la   \cos\theta_{1,2} \cos\theta_{2,k}\ra.
\end{equation}
Neglecting correlations between successive scattering angles, we find that 
\begin{equation}
\label{rkfac2}
\la    \cos\theta_{1k}\ra
\approx
 \la \cos\theta_{1,2}\ra \la  \cos\theta_{2,k}\ra.
\end{equation}
Repeating this procedure for $\la \cos\theta_{2,k}\ra$, then for $\la \cos\theta_{3,k}\ra$, and so on, we find that 
\begin{equation}
\label{rkfac}
\la   \cos\theta_{1k}\ra
\approx
 \prod_{i=1}^{k-1}   \la  \cos\theta_{i,i+1}\ra,    \qquad k\ge 2.
\end{equation}
Although the  granular temperature decreases in time due to the inelastic character of collisions, $\langle \cos\theta_{i,i+1}\rangle$ is independent of $i$ (i.e.,  it does not depend on time) because the  scattering angle in a collision depends on the normal restitution coefficient, not the speed of the colliding particles.
Therefore,
\begin{equation}
\label{rcprod}
\la  \cos\theta_{1k}\ra
\approx
  \la  \cos\theta_{1,2}\ra^{k-1}.
\end{equation}
 Note that all the arguments shown so far are analogous to those for molecular gases \cite{Yuste2024}.

We now relate the computation of $\la \cos\theta_{1,2}\ra$ to the mean-persistence ratio  $\la \omega \ra$. This quantity is usually defined in kinetic gas theory as (see Sec.~5.5 of chapter 5 of \cite{CCC70}, for example):
\begin{equation}
\label{omega}
\la \omega \ra=\la\frac{v_2}{v_1} \cos\theta_{1,2}\ra.
\end{equation}
In order to connect $\la \cos\theta_{1,2}\ra$ with $\la \omega \ra$   we write
\begin{align}
\label{rrome}
\la \cos\theta_{1,2}\ra=&
\la\frac{v_1}{v_2  } \frac{v_2  }{v_1  } \cos\theta_{1,2}\ra
\approx \la \frac{v_1}{v_2  }\ra \; \la\frac{v_2}{v_1} \cos\theta_{1,2}\ra\equiv \Omega,
\end{align}
where
\begin{equation}
\label{OmegaDef}
\Omega=\la\frac{v_1}{ v_2}\ra \la \omega \ra.
\end{equation}
\sby{Texto para responder a punto 3 del Reviewer 3 }
\vicente{For the approximation in Eq.~\eqref{rrome} to hold, we  assumed that the fluctuations of    $\cos\theta_{12}$ and $v_1/v_2$ around their mean values are small (with the mean value of $v_1/v_2$ close to 1) and that  $\cos\theta_{12}$ and $v_1/v_2$  are weakly correlated.}
We will refer to \vicente{$\Omega$}
as perseverance.
It is closely related to the collision mean-persistence ratio $\la \omega \ra$ and equal to it when $\la{v_1}/{v_2}\ra=1$ (elastic collisions).

Combining Eqs.~\eqref{r1rkAb}, \eqref{rcprod}, and \eqref{rrome}, we find (see Eq.\ \eqref{ckge2})
\begin{equation}
	\label{rbrkok}
 c_k\equiv \frac{2\la  \rb_1\cdot \rb_{k}\ra}{\la r^2\ra}  \approx    \Omega^{k-1} 
\end{equation}
for $k\ge 2$ (recall that $c_1\equiv 1$).
This is one of the key results of the present article. 
Substituting this result into Eq.~\eqref{MSDles2} yields
\begin{equation}
\label{MSDRWNfinita}
\la R^2_N\ra
\approx  \la r^2\ra \sum_{k=0}^{N-1} (N-k) \Omega^k
=\frac{\la r^2\ra}{1-\Omega}\left[N-\frac{\Omega\left(1-\Omega^N\right)}{1-\Omega}\right]
\end{equation}
For large $N$
\begin{equation}
\label{MSDRWN}
\la R^2_N\ra
\approx   \frac{\la r^2\ra}{1-\Omega}\, N
\end{equation}
or, in terms of the time,
\begin{equation}
\label{MSDRW}
\la R^2(t)\ra= \la N(t)\ra \,\frac{\la r^2\ra}{1-\Omega}.
\end{equation}
In terms of the effective free path, this relation is equivalent to 
\begin{equation}
\label{ele2pa}
\frac{\ell_e^2}{\la r^2\ra}=\frac{\la R^2\ra}{\la N \ra \la r^2\ra}\approx \frac{1}{1-\Omega}.
\end{equation}

Equation \eqref{MSDRW} (or Eq.~\eqref{ele2pa}) is  a fundamental equation in this work. It tells us how the MSD of the  tracer (or intruder) particle differs from that of a particle scattered by collisions with equiprobable spatial re-emission. The ``elementary'' relation $\la R^2\ra= \la N \ra \la r^2\ra$ (equivalent to setting $\Omega=0$) would apply in that case. Equation \eqref{MSDRW} shows us that the particle's persistence in maintaining its direction after collisions results in a positively correlated random walk, resulting in an actual MSD exceeding that without persistence.

\subsection{M. Smoluchowski and J. H. Jeans}

The stochastic, random walk procedure, where we follow the particle of interest (intruder) and treat each collision as a random event, comes from M. Smoluchowski, a pioneer in what we now call stochastic processes. This is the approach we have used here.
Although there are some differences between our method and the method of  Smoluchowski, there are also strong similarities (see Sec.~11 of Ref.\ \cite{Smoluchowski1906b}). Smoluchowski assumed a constant scattering angle $\theta$  (which he denotes as $\varepsilon$), and a fixed free path length $l$ ($l\equiv r_1=r_2=\ldots=\text{const.}$).  A recurrence relation between $\la R^2_N \ra$ and $\la R^2_\vicente{N-1} \ra$ was then found, yielding exactly Eq.~\eqref{MSDRW} for large $N$, if we equate $\Omega$ with $\cos \varepsilon$ and $\la r^2\ra$ with $2l^2$.

On the other hand, J. H. Jeans knew (see Sec.~168 of Ref.\ \cite{Jeans1940}) that collision persistence had to be considered for a more accurate description of molecular diffusion in gases. Smoluchowski was also aware of this: see Sec.~7 of Ref.\ \cite{Smoluchowski1906a}, where Jeans' article \cite{Jeans1904}  is referenced. Taking collision persistence into account, Jeans suggested the improved formula $D=D_\text{eKT}/(1-\omega)$ (for elastic collisions only), identifying $\omega$ as the mean-persistence ratio. However, Jeans did not connect his formula with Smoluchowski's, i.e., the parameter $\omega$ (which he denotes as $\theta$) is an estimate of Smoluchowski's $\cos \varepsilon$.

Based on the preceding discussion, it might be appropriate to refer to our equation \eqref{MSDRW} as the  inelastic Smoluchowski-Jeans equation.

\subsection{Connection to Polymer Physics}
\label{sec:Polymer}

The trajectory of a particle (the intruder) after $N$ collisions, $\{\rb_1,\rb_2,\ldots, \rb_N\}$, can be viewed as a polymer configuration with $N+1$ monomers, where $\rb_i$ is the vector connecting monomer $i$ to $i-1$ (monomers are placed at intruder collision points).
From this viewpoint, readers familiar with polymer physics will recognize our effective free path $\ell_e$, defined by Eq.~\eqref{MSDle3}, as the Kuhn length \cite{Rubinstein2003}.
Furthermore, if we identify $\la r^2\ra$ with the square of the bond length, then $\la R^2\ra/(N \la r^2\ra)$ (see Eq.\eqref{ell2redSer}) is the Flory characteristic ratio in polymer physics.

From equation \eqref{rbrkok}, we learn that the correlation between collisions  decay exponentially: 
\begin{equation}
\label{descorre}
\la \rb_1\cdot \rb_{1+n}\ra   \sim    \Omega^{n} =\exp(n \ln\Omega)=\exp(-n/\tilde{n} )
\end{equation}
where
\begin{equation}
\label{tilden}
\tilde n=-1/\ln\Omega 
\end{equation}
are the persistence collisions.
Borrowing the terminology employed to define persistence length in polymers (see Ref.~\cite{Grosberg1994}, Sec. 2.3), 
persistence collisions $\tilde n$ can \emph{roughly} be considered  as the maximun number of collisions that  the intruder can have while its trajectory $\{\rb_1,\rb_2,\ldots, \rb_{\tilde n}\}$ remains roughly straight; at greater number of collisions, the bending of the $\rb_k$ connections destroy the memory of the trajectory direction. 
A more detailed discussion on persistence collisions for elastic collisions is given at the end of section IV of Ref.\ \cite{Yuste2024}.

\subsection{MSD regimes with respec to $N$ and time}
\label{sec:MSDregimes}
Although this article primarily focus on intruder diffusion in the long-time (or high number of collisions)  regime (where Equations \eqref{MSDRWN} and \eqref{MSDRW} hold), in this  subsection we provide a brief qualitative overview of the MSD for other time regimes.

\subsubsection{Scaling of the MSD with the number of col\vicente{l}isions}
\label{sec:ScaRN}
In the preceding subsection \ref{sec:Polymer}, we established that the trajectory approximates a straight line when the number of collisions $N$ is less than or of the order of the persistence collisions  $\tilde{n}$. Consequently, for $N\lesssim \tilde{n}$, the displacement $R$ scales linearly with $N$, leading to  $\langle R^2\rangle\sim N^2$.  This outcome can also be deduced directly from  Eq.~\eqref{MSDRWNfinita}: the condition for a quasi-ballistic trajectory over a substantial number of collisions requires  $\Omega$ approaching unity, as then $\tilde n=-1/\ln\Omega\approx 1/(1-\Omega)$  becomes relatively large.  In this case, Eq.~\eqref{MSDRWNfinita} becomes  
\begin{equation}
	\label{Nsmall}
	\frac{\la R^2_N\ra}{\la r^2\ra}=\frac{N+N^2}{2}+\frac{N-N^3}{6}\left(1-\Omega\right)
	+\mathcal{O}\left(1-\Omega\right)^2.
\end{equation}
This equation tell us that  the MSD is ballistic, i.e., $\la R^2_N\ra$ grows as $N^2$,  for $1\ll N\ll \tilde n$. 
This ballistic regime ceases when the $N^2$ term approaches the magnitude of the $N^3 (1-\Omega)$ term, a condition satisfied when $N\approx 1/(1-\Omega)\approx \tilde n$. Consequently, for $N\gg \tilde n$, the ballistic displacement regime vanishes (as the bending of the $\rb_k$ connections disrupts trajectory direction memory), and the diffusive regime, $\la R^2_N\ra \sim N$, as described by Eq.~\eqref{MSDRWN}, is established.

\subsubsection{Scaling of the the average number of collisions with time}
\label{sec:ScaNt}

From Eq.~\eqref{s0t}, it is easy to see that   $s_0\sim \ln(t/t_\zeta)$ for $t\gg t_\zeta  $ while $s_0\sim t/\tau_0(0)$ for $t\ll t_\zeta$.
The physical meaning of this expression is clear.   Equation \eqref{HaffLaw0} shows that the intruder's temperature (i.e., its velocity) changes negligibly for $t\ll t_\zeta$.
This occurs because energy loss from collisions with gas particles is minimal in this regime, keeping the intruder's velocity close to its initial value.  Therefore,  the number of collisions $s_0(t)$   is simply the time over the mean collision time:  $t/\tau_0(0)$.  
At times much greater than $t_\zeta$,  the  intruder's velocity and collision frequency have changed substantially from their initial values and  both decay as $ t_\zeta/t$. This results in a logarithmic growth  $s_0\sim \ln(t/t_\zeta) $ of the number of collisions  for $t\gg t_\zeta$.

\subsubsection{Scaling of the MSD with time}
\label{sec:ScaRt}

To determine the detailed time dependence of the MSD for all times we should use Eq.~\eqref{MSDRWNfinita} together with Eq.~\eqref{RTRN}.  
However, a simpler analysis is sufficient for our  purpose of determining the  scaling of the MSD with time.
Thus, we assume that the number of collisions $N$ up to a given $t$ is similar to its mean value,  $N\approx \la N\ra \equiv s_0(t)$ (i.e., we assume that the variance of $N$ is small). This is equivalent to the approximation $\xi_N(t)\approx \delta(N-s_0(t))$ in equation \eqref{RTRN}, where $\delta$ is the Dirac delta.
In other words, to estimate the MSD for all times we will use the relationship
\begin{equation}
	\label{x}
	\la R^2(t)\ra\approx  \la R^2_{N=s_0(t)} \ra
\end{equation}
with  $\la R^2_{N} \ra$  given by Eq.~\eqref{MSDRWNfinita}.

Obviously,   for times less than or comparable to the mean collision time $\tau_0(0)$, i.e., before the first collision, the  trajectory of the intruder is straight, and thus  $\la R^2(t)\ra\sim t^2$.
We will call this (trivial) temporal regime the pre-collisional regime.  
But we know that, due to the persistence of the collisions, the trajectory of the intruder remains roughly straight  beyond the first collision for $\tilde n$ collisions, i,e.,  until times of order of $t_{\tilde{n}}$ where  $s_0(t_{\tilde{n}})=\tilde{n}$.
Thus, beyond the precollisional regime,  we distinguish three temporal regimes depending on the values of $t_{\tilde{n}}$ and $t_\zeta$:

\begin{enumerate}
	
	\item \textit{Early time regime or ballistic regime:}       $\tau_0(0)\ll t\ll \text{min}\{ t_{\tilde{n}},t_\zeta\}$. In this case,  (i) the movement of the intruder is roughly straight, $\langle R^2\rangle\sim s_0^2$, because $t\ll t_{\tilde{n}}$ and  (ii) $s_0(t)\sim t$ because $t\ll t_\zeta$. These two facts imply ballistic motion, $\langle R^2\rangle\sim s_0^2\sim t^2 $.

	\item \textit{ Intermediate time regime.}  We distinguish two cases:
	
	\begin{enumerate}
		
		\item  $ t_{\tilde{n}}\ll t \ll t_\zeta$. Here the motion of the intruder is diffusive in the number of collisions,    $\langle R^2\rangle\sim s_0$, because $t_{\tilde{n}}\ll t$. Since $s_0(t)\sim t$ because $t \ll t_\zeta$, this implies  $\langle R^2\rangle\sim s_0 \sim t $, i.e.,  normal diffusive behavior. Note that this intermediate normal diffusive regime becomes the  \textit{final} diffusive regime for elastic collisions because, in this case,  $t_\zeta\to\infty$ (see Eq.~\eqref{tzetatau0}). This normal diffusive regime has been studied in detail using our random walk approach in Ref.~\cite{Yuste2024}. 
		
		\item  $ t_\zeta \ll t \ll t_{\tilde{n}}$. Here the motion of the intruder is ballistic with respect to the number of collisions,  $\langle R^2\rangle\sim s_0^2$, because $t\ll t_{\tilde{n}}$. But, on the other hand, $s_0(t)\sim \ln t$ when $t_\zeta\ll t$. Thus,  $\langle R^2\rangle\sim s_0^2 \sim\ln^2 t $.
		Note that  $ t_\zeta \ll  t_{\tilde{n}}$ holds when $\alpha$ is not close to 1 (see  Eq.~\eqref{tzetatau0}) and $\Omega$ is very close to 1 (see  Eq.~\eqref{tilden}).  In Sec.~\ref{sec:Omega} we will show that $\Omega$ becomes closer to  1 as  $m_0/m$ increases and $\sigma_0/\sigma$ decreases.
		
	\end{enumerate}  
	
	\item \textit{Long-time regime:}  $ \text{max}\{ t_{\tilde{n}},t_\zeta\}\ll t$.	Here, again, the motion is diffusive with respect to the number of collisions, $\langle R^2\rangle\sim s_0$, because $t_{\tilde{n}}\ll t$. But now the number of collisions decays logarithmically,   $s_0 \sim\ln t $, because $t_\zeta \ll t$. This implies  $\langle R^2\rangle\sim s_0 \sim\ln t $, which is a subdiffusive behaviour  (sometimes termed ultraslow as the MSD grows slower than any power of time).    
\end{enumerate}

The intermediate regime can not be detected when $t_{\tilde{n}}$ and $t_\zeta$ are similar. Furthermore, if these times are not much larger than $\tau_0(0)$, the time dependence of the MSD does not exhibit the well-defined functional forms of the first two regimes, and only the long time subdiffusive regime is clearly observable (aside from the pre-collisional regime at very short times).
This long-time temporal regime is main subject of this article, and it is the regime to which Eqs.~\eqref{MSDRWN} and \eqref{MSDRW} apply.
In Refs.~\cite{Bodrova2015,Bodrova2024,Bodrova2025},  Bodrova et al.   considered the all-time temporal evolution of the MSD for an intruder in a granular gas in the HCS state (and also in granular gas mixtures). They provided an equation for the MSD at any instant and described two temporal regimes: ballistic for $t\ll t_\zeta$ and  subdiffusive  (logarithmic, ultraslow)  for $t_\zeta\ll t$, although no intermediate temporal regime was identified. 

Although the mechanisms and regimes identified differ, the presence of various regimes in the MSD has been previously observed in MD simulations of freely cooling viscoelastic grains \cite{Bodrova2012} and confined granular gases \cite{GGGBS24}.

\section{Perseverance   for inelastic hard spheres}
\label{sec:Omega}

Our main formula \eqref{MSDRW}, or \eqref{ele2pa}, tells us that collision persistence modifies the MSD of a particle by a factor $1/(1-\Omega)$ relative to the MSD without persistence. To make this equation useful, we need to determine the perseverance $\Omega=\la v_1/v_2\ra \omega$ in terms of the parameter space of the system (intruder immersed in a granular gas in the HCS). Since the mean-persistence ratio $\la \omega \ra$ is a dimensionless quantity, it should be a function of the mass $m_0/m$ and diameter $\sigma_0/\sigma$ ratios and the coefficients of restitution $\alpha$ and $\alpha_0$. 
In the following two subsections, we will provide expressions for the mean-persistence ratio $\la \omega \ra$ and the average pre- to post-collisional intruder velocity ratio, $\la v_1/v_2\ra$, for this system.

\subsection{Mean-persistence ratio for inelastic hard spheres collisions}
\label{sec:persis}

The mean-persistence ratio for elastic hard sphere collisions has been exactly calculated in Sec.~5.5 of Ref.~\cite{CCC70}. However, for inelastic collisions, the evaluation of $\la \omega \ra$ requires to take some approximations since the exact form of the velocity distributions of the intruder $f_0$ and granular gas particles $f$ are not known to date \cite{GarzoBook19}. As usual, to estimate $\la \omega \ra$ for inelastic hard spheres we take Maxwellian distributions defined at their corresponding temperatures for $f_0$ and $f$.    
Some technical details on this calculation are provided in the Appendix \ref{sec:AppA}. The final expression of $\la \omega \ra$ for a three-dimensional system ($d=3$) is 
\begin{equation}
\label{omegral}
 \la \omega \ra  =  1-\frac{1}{2}\, \frac{1+\alpha_0}{1+m_0/m} [1-\tilde \omega(\beta)],
\end{equation}
with
\begin{equation}
\tilde \omega(x)=\frac{x}{2}\left[ \frac{x}{\sqrt{1+x}} \ln\left( \frac{1+\sqrt{1+x}}{\sqrt{x}} \right) -1\right].
\end{equation}
\sby{texto respuesta a punto 4 de Reviewer 3:}
\vicente{Note that the dependence of $\langle \omega\rangle$ on the restitution coefficient $\alpha$ of the granular gas comes from the dependence of $\beta$  on $\alpha$  (recall  Eq.~\eqref{ecubeta}).
}
If we set $\alpha=\alpha_0=1$, then $T_0/T=1$ and $\beta=m/m_0$, so that we get back the standard result for elastic hard spheres \cite{CCC70}.

\subsection{Average Pre- to Post-Collisional Intruder Velocity Ratio}
\label{sec:veloci}

We now evaluate the average $\la v_1/v_2\ra$, where $v_1$ and $v_2$ are the modulus of the  pre- and post-collisional velocities of the intruder. Given that this calculation is quite intricate, we approximate this as
\begin{equation}
	\label{v1v2aprox}
\la \frac{v_1}{v_2} \ra \approx \frac{\bar{v}_0(t)}{\bar{v}_0(t+\tau_0)}
\end{equation}
where $\tau_0(t)=1/\nu_0(t)$ is the mean time  between two successive collisions of the intruder with the gas particles, and $\bar{v}_0(t)$ denotes the average speed modulus of the  intruder at time $t$.

To estimate $\bar{v}_0(t)$, we consider the time evolution of the intruder's temperature $T_0(t)$. 
From Eqs.~\eqref{2.1} and \eqref{2.2}  one realizes that 
\beq
\label{5.1}
\partial_t \ln T_0=-\zeta(t).
\eeq
Because  $T_0\propto \bar{v}_0^2$,  Eq.~\eqref{5.1} implies 
\begin{equation}
\label{5.2}
\frac{d \bar{v}_0}{dt}=-\frac{1}{2}\zeta(t)\bar{v}_0.
\end{equation}
So, if we integrate Eq.\ \eqref{5.2} between $t$ and $t+\tau_0$, we get
\begin{align}
	\bar{v}_0(t+\tau_0)-\bar{v}_0(t)=&-\int_{t}^{t+\tau_0}\frac{\zeta(t')}{2}\,   \bar{v}_0(t') dt'\nonumber\\
	=&	-\frac{\zeta(t)\tau_0(t)}{4} \left[\bar{v}_0(t+\tau_0)+\bar{v}_0(t)\right].
\label{v0e1}
\end{align}
In the last expression, we (i)  have assumed that the cooling rate $\zeta(t)$ does not change much during the microscopic time interval $\tau_0$ and, so, it can be considered constant in the integral, 
and (ii) we have approximated the remaining integral using the trapezoidal rule. 
From Eq.~\eqref{v0e1} we find
\begin{equation}
\label{v0e2}
\frac{\bar{v}_0(t)}{\bar{v}_0(t+\tau_0)}=\frac{1+\zeta\tau_0/4}{1-\zeta \tau_0/4}.
\end{equation}
Since $\zeta(t)\propto T(t)^{1/2}$ and $\tau_0(t)=\nu_0(t)^{-1}\propto T(t)^{-1/2}$, it turns out that  $\zeta(t)\tau_0(t)=\zeta(t)/\nu_0(t)$ does not depend on time. Thus, from Eq.\ \eqref{nu0nu} one achieves the expression
\begin{equation}
	\label{zetanu0}
	\zeta(t)\tau_0(t)= \frac{1-\alpha^2}{d\Upsilon},
\end{equation}
where use has been made of the result \cite{Abad2022}
\begin{equation}
\zeta(t)\tau(t)= \frac{1-\alpha^2}{d}.
\end{equation}
Inserting Eq.~\eqref{zetanu0} into Eq.~\eqref{v0e2}, and taking into account Eq.~\eqref{v1v2aprox}, we find 
\begin{equation}
\label{vpvA}
\left\langle \frac{v_1}{v_2}\right\rangle  \approx
\frac{4d\Upsilon-\alpha^2+1 }{4d\Upsilon+\alpha^2-1 }.
\end{equation}
Finally, from Eqs~\eqref{OmegaDef}, \eqref{omegral}, and \eqref{vpvA}, we obtain the following approximated expression for the perseverance $\Omega$:
\begin{align}
\label{Omegafinal}
\Omega
 &=\frac{4d\Upsilon+1-\alpha^2}{4d\Upsilon-1+\alpha^2} \left\{1-\frac{1}{2}\, \frac{1+\alpha_0}{1+m_0/m}\Big[1-\tilde{\omega}(\beta)\Big]\right\}.
\end{align}
This equation, along with Eq.~\eqref{rbrkok}  (and then Eqs.~\eqref{MSDRWNfinita} and \eqref{MSDRW}), is  likely the main result of the present work.

\sby{ Parte de la respuesta a Reviewer 2, punto 4 . El resto de la respuesta es el apéndice B:}
\vicente{Equation \eqref{v0e1} was obtained by approximating the integral using the trapezoidal rule. One might  wonder how our estimates for the velocity ratio, Eq.~\eqref{v0e2}, and consequently $\Omega$ in Eq.~\eqref{Omegafinal}, would change if we had used a different approximation for the integral,  such as the rectangle rule, instead of the trapezoidal rule. One expects the difference between using one rule or the other to be small because  $\tau_0$ is, in general,  small. This is indeed what happens, as shown in Appendix \ref{App:IntRules}. In any case, the trapezoidal rule is generally more accurate than rectangular rules, and it is the one we have chosen to use in this paper.
}

\section{ Dependence of $\la \rb_1\cdot \rb_k\ra$ on $k$ and MSD  in units of $\la r^2 \ra$.  Comparison with simulations}
\label{sec:MSDsimu}

In Sec.~\ref{sec:SmoJeans}, we derived the key result \eqref{rbrkok}, showing that $\la \rb_1\cdot \rb_k\ra$ is proportional to $\Omega^{k-1}$. This makes the reduced collisional series \eqref{ell2redSer} a geometric series with ratio $\Omega$. From this, we got our main result:  Eqs.~\eqref{MSDRW} and \eqref{Omegafinal}   for the MSD of the intruder. 
Now, we want to assess the accuracy of these theoretical predictions. In this section we consider the diffusion of an intruder in a 3D granular gas.
We will test the validity of Eqs.~\eqref{rbrkok}, \eqref{MSDRW}, and our expression for  $\Omega$, Eq.~\eqref{Omegafinal},  comparing them with DSMC results.

In

\subsection{Dependence of $\la \rb_1\cdot \rb_k\ra$ on $k$}

From the definitions of $c_k$ in Eq.~\eqref{ckDef}, we see that studying the $k$-dependence of $\la \rb_1\cdot \rb_k\ra$ is equivalent to studying the $k$-dependence of $c_k$. Equation~\eqref{rbrkok} implies
\begin{subequations}
	\label{ckOmega}
	\begin{align}
		\frac{c_{k+1}}{c_k} \approx \Omega, \quad \forall k
	\end{align}
\end{subequations}
i.e., $c_{k+1}/c_k\approx \Omega$ for all $k$.
To validate these relations, we will compare  the theoretical expression of  $\Omega$ (given by Eq.~\eqref{Omegafinal}) with  the numerical values of  $c_{k+1}/c_k$  
obtained by means of the DSMC method \cite{Bird94} for inelastic hard spheres (implementation details of this method are provided in the Appendix \ref{sec:AppB}).

Given that the parameter space of our system (constituted by the parameters $\alpha_0$, $\alpha$, $m_0/m$, $\sigma_0/\sigma$, and the reduced density $n \sigma^3$) is quite large, a full comparison  between theory and simulations  exceeds the scope of this article. For this reason, we will compare results for a limited set of representative cases.  Additionally, we consider the low-density regime ($n \sigma^3 \to 0$), and hence the pair correlation functions are $\chi=\chi_0=1$.

In Table \ref{tabla1}, simulation results are presented for the initial four values of $c_{k+1}/c_k$ for 24 cases with different masses, sizes, and normal restitution coefficients. The agreement between simulation and theory is excellent: simulated values for $k=1$, 2, 3, and 4 closely match the theoretical values of $\Omega$ obtained from Eq.~\eqref{Omegafinal}; differences are rarely above a couple of hundredths and usually much smaller. This is especially true for $c_{2}/c_1$, where the difference from the theoretical $\Omega$ is always less than one hundredth. This is of significant relevance given that the first two terms in the reduced collisional series are the most important (especially as $\Omega$ decreases).

\begin{table*}[th]
	\centering
	\begin{tabular}{cccccccccc}
		\toprule
	\;\; $\alpha$\;\; & $\;\;\alpha_0\;\;$ & $\frac{m_0}{m}$ & $\frac{\sigma_0}{\sigma}$ &  $c_2/c_1$& $c_3/c_2$ &  $c_4/c_3$ & $c_5/c_4$  & $\Omega$ \\
		\hline
0.4 & 0.4 & 1. & 1. & 0.683(0) & 0.683(1) & 0.683(2) & 0.684(2) & 0.6720 \\
0.6 & 0.6 & 1. & 1. & 0.591(0) & 0.595(1) & 0.597(2) & 0.598(3) & 0.5838 \\
0.8 & 0.8 & 1. & 1. & 0.500(0) & 0.506(1) & 0.510(2) & 0.513(3) & 0.4940 \\
1.  & 1.  & 1. & 1. & 0.413(0) & 0.421(1) & 0.425(2) & 0.428(5) & 0.4058 \\
		\hline
 0.4 & 0.4 & 2. & 1. & 0.842(0) & 0.840(1) & 0.840(1) & 0.839(1) & 0.8331 \\
0.6 & 0.6 & 2. & 1. & 0.767(0) & 0.767(1) & 0.766(1) & 0.766(2) & 0.7585 \\
0.8 & 0.8 & 2. & 1. & 0.681(0) & 0.683(1) & 0.684(1) & 0.684(2) & 0.6749 \\
1.0 & 1.0 & 2. & 1. & 0.591(0) & 0.596(1) & 0.598(1) & 0.599(2) & 0.5868 \\
		\hline
0.4 & 0.4 & 0.5 & 1. & 0.527(0) & 0.533(1) & 0.536(2) & 0.538(3) & 0.5209 \\
0.6 & 0.6 & 0.5 & 1. & 0.428(0) & 0.436(1) & 0.442(2) & 0.446(4) & 0.4241 \\
0.8 & 0.8 & 0.5 & 1. & 0.337(0) & 0.346(1) & 0.352(3) & 0.356(8) & 0.3313 \\
1.0 & 1.0 & 0.5 & 1. & 0.256(0) & 0.262(1) & 0.268(4) & 0.274(16) & 0.2447 \\
		\hline
 0.4 & 0.4 & 1. & 2. & 0.624(0) & 0.627(1) & 0.627(2) & 0.629(2) & 0.6143 \\
0.6 & 0.6 & 1. & 2. & 0.551(1) & 0.556(1) & 0.558(1) & 0.560(2) & 0.5439 \\
0.8 & 0.8 & 1. & 2. & 0.480(0) & 0.487(1) & 0.491(2) & 0.493(4) & 0.4740 \\
1.0 & 1.0 & 1. & 2. & 0.413(0) & 0.421(1) & 0.425(2) & 0.429(4) & 0.4058 \\
		\hline
 0.4 & 0.4 & 1. & 0.5 & 0.748(0) & 0.746(1) & 0.746(1) & 0.745(1) & 0.7420 \\
0.6 & 0.6 & 1. & 0.5 & 0.641(0) & 0.643(1) & 0.644(1) & 0.645(2) & 0.6371 \\
0.8 & 0.8 & 1. & 0.5 & 0.527(1) & 0.533(1) & 0.536(1) & 0.540(3) & 0.5220 \\
1.0 & 1.0 & 1. & 0.5 & 0.413(0) & 0.421(1) & 0.425(2) & 0.429(3) & 0.4058 \\
		\hline
 0.4 & 0.7 & 1. & 1. & 0.581(0) & 0.585(1) & 0.586(1) & 0.588(2) & 0.5735 \\
0.6 & 0.7 & 1. & 1. & 0.559(0) & 0.564(1) & 0.567(1) & 0.569(2) & 0.5526 \\
0.8 & 0.7 & 1. & 1. & 0.529(0) & 0.534(1) & 0.538(2) & 0.540(3) & 0.5229 \\
1.0 & 0.7 & 1. & 1. & 0.491(0) & 0.498(0) & 0.501(2) & 0.504(3) & 0.4849 \\
\toprule
	\end{tabular}
	\caption{
DSMC  values of the first four ratios $c_{k+1}/c_k$ ($k=1, 2, 3, 4$) of the reduced collisional series \eqref{ell2redSer} for several values of $\alpha$, $\alpha_0$, $m_0/m$, and $\sigma_0/\sigma$. The last column shows $\Omega$ from Eq.~\eqref{Omegafinal}. Simulation values and standard errors are rounded to the nearest thousandth. A standard error less than $5 \times 10^{-4}$ is denoted by (0).	
	}.
	\label{tabla1}
\end{table*}

\begin{figure}
	\begin{center}
		\includegraphics[width=0.98\columnwidth]{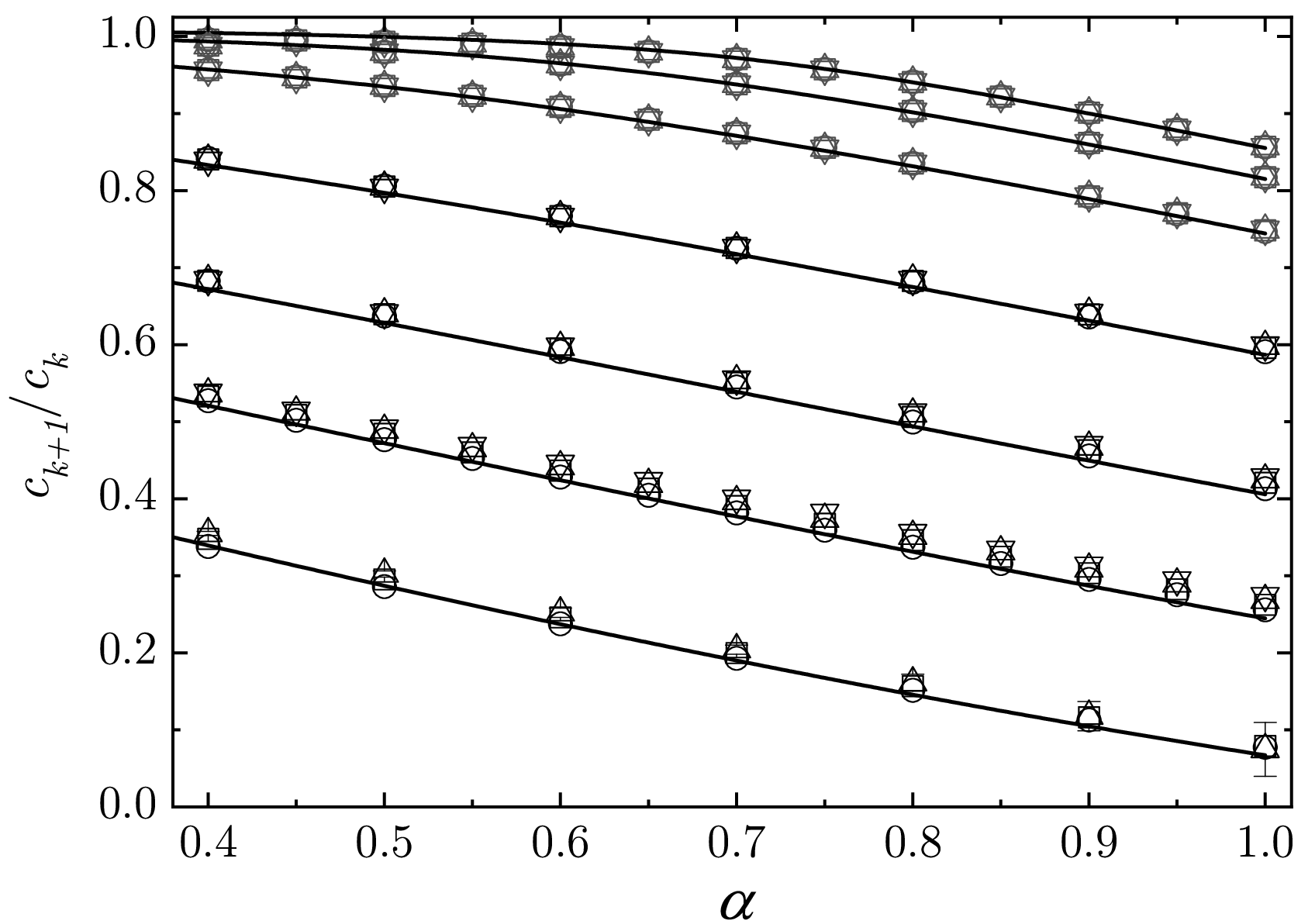}
	\end{center}
\caption{DSMC values of the first ratios $c_{k+1}/c_k$ of the reduced collisional series vs the (common) coefficient of restitution $\alpha=\alpha_0$ for $k=1, 2, 3, 4$ (circles, squares, up triangles, down triangles, respectively) and, for from top to bottom, $m_0/m=8, 6, 4, 2, 1, 1/2, 1/8$ with $\sigma_0/\sigma=1$.
Here, and in the rest of the figures, the absence of error bars indicates that the standard error is smaller than the size of the symbols. 
Due to large statistical errors, especially at $\alpha=0.9$ and 1, we have not included simulation data for $c_5/c_4$ when $m_0/m=1/8$. The error bars shown for $m_0/m=1/8$ correspond to the uncertainty in $c_4/c_3$.
The solid lines represent $\Omega$ defined by  Eq.~\eqref{Omegafinal}.
\label{fig:ck1ckA}}
\end{figure}

Figure \ref{fig:ck1ckA} further compares simulated $c_{k+1}/c_k$ values for $k=1$, 2, 3, and 4 with the theoretical values of $\Omega$ for additional cases. The figure shows that the agreement between simulation and $\Omega$ worsens as $k$ increases, though this effect diminishes with increasing the mass ratio $m_0/m$. For $m_0/m\ge 2$, simulation symbols for different $k$ values are barely distinguishable.
For a given value of $k$, $c_{k+1}/c_k$ increases with $m_0/m$, indicating that the correlation between the initial displacement $\rb_{1}$ and the $k$-th displacement $\rb_{k}$ decreases more slowly when $m_0/m$ gets bigger. This makes sense: if the intruder is much heavier than the granular gas particles, collisions will hardly perturb the intruder's trajectory, so the direction after $k$ collisions (if $k$ is not too large) will remain close to the initial $\rb_{1}$.
This effect becomes stronger as the collisions become more inelastic (as $\alpha_0=\alpha$ becomes smaller) because the normal component of the relative velocity along the line of separation between the centers of the colliding sphere decreases. In other words, inelastic particles bounce less than elastic ones, so they tend to keep going in the same direction as before the collision (this effect can be seen in the animations of Ref.~\cite{Santos2008}).

\begin{figure}
	\begin{center}
		\includegraphics[width=.98\columnwidth]{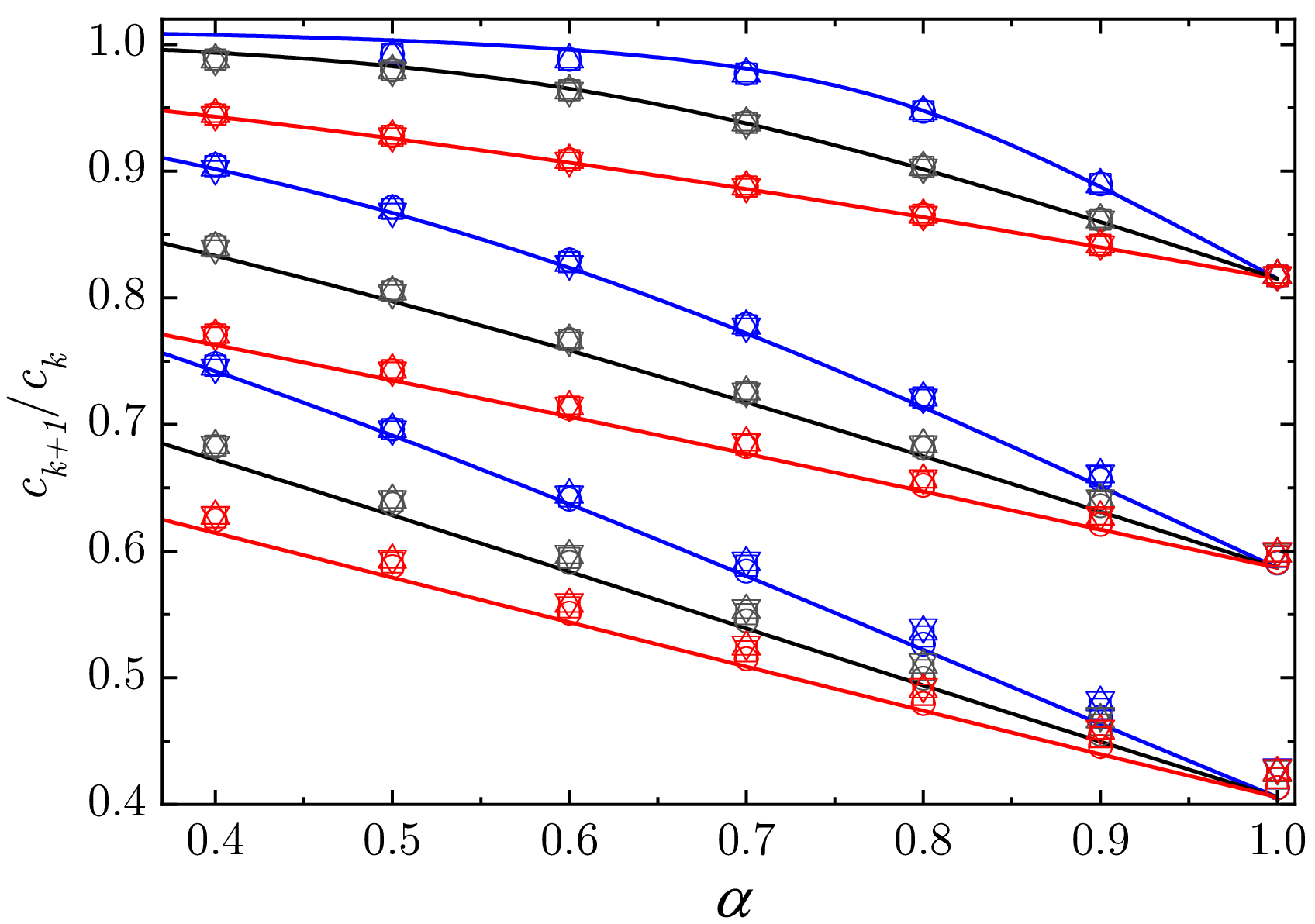}
	\end{center}
\caption{\vicente{(Color online.)} DSMC values of $c_{k+1}/c_k$ vs the (common) coefficient of restitution $\alpha=\alpha_0$ for $k=1, 2, 3, 4$ (circles, squares, up triangles, down triangles, respectively) and, for from top to bottom, $\{m_0/m, \sigma_0/\sigma\}=\{6, 1/2\}$, $\{6, 1\}$, $\{6, 2\}$, $\{2, 1/2\}$, $\{2, 1\}$, $\{2, 2\}$, $\{1, 1/2\}$, $\{1, 1\}$  and $\{1,2\}$.  The solid lines represent $\Omega$ defined by  Eq.~\eqref{Omegafinal}. 
	\vicente{The \{blue, black, red\} color is used for $\sigma_0/\sigma=\{1/2,1,2\}$}.
\label{fig:ck1ckB}}
\end{figure}

In Fig.~\ref{fig:ck1ckB}, we also compare simulated $c_{k+1}/c_k$ values for $k=1$, 2, 3, and 4 with the theoretical $\Omega$ for three mass ratios ($m_0/m=6$, 2, and 1), and three size ratios ($\sigma_0/\sigma=1/2$, 1, and 2). Here, the simulation values obtained from the DSMC method again agree well with the theoretical prediction for $\Omega$. We see that for given values of the mass ratio $m_0/m$ and the (common) coefficient of restitution ($\alpha=\alpha_0$), $\Omega\sim \la \rb_1\cdot \rb_{1+k}\ra^{1/k}$ increases when $\sigma_0/\sigma$ decreases, meaning collision correlation decreases more slowly when $\sigma_0/\sigma$ is smaller. This, along with the $m_0/m$ dependence we find in Fig.~\ref{fig:ck1ckB}, shows that collision correlations decrease more slowly ($\Omega$ is larger) when $m_0/m$ is higher, $\sigma_0/\sigma$ is lower, and collisions are more inelastic.

Figures \ref{fig:ck1ckA} and \ref{fig:ck1ckB} show that the dependence of $c_{k+1}/c_k$ on  the coefficients of restitution  is approximately linear, provided $m_0/m$ is not too large and $\sigma_0/\sigma$ is not too small. We also notice that curves for different $m_0/m$ values but the same $\sigma_0/\sigma$ are roughly parallel. However, when $m_0/m$ is large and $\sigma_0/\sigma$ is small, the lines curve and tend to run parallel to the line $c_{k+1}/c_k=1$.
This nonlinear behavior is evident in the case $m_0/m=6$, especially when $\sigma_0/\sigma=1/2$. This case also shows that when $\alpha_0\lesssim 0.55$, the simulated $c_{k+1}/c_k$ values are so close to 1 that $\Omega$ becomes unphysical, exceeding 1. This leads to unphysical (negative) MSD values according to Eq.~\eqref{MSDRW}. However, Eq.~\eqref{MSDRW} correctly predicts a very large MSD when $c_{k+1}/c_k$ values are close to 1, due to slow convergence of the reduced collisional series. This makes sense physically, as an intruder with $\Omega\approx 1$ acts like a quasi-ballistic particle for many collisions. Remember (from Eq.~\eqref{tilden}) that the persistence collisions (the maximum number of collisions for which the trajectory can be considered ballistic) are given by $\tilde n=-1/\ln\Omega$.

Finally, Fig.~\ref{fig:ck1ckD} 
shows $c_{k+1}/c_k$ versus $\alpha_0$  for a fixed value of $\alpha$ ($\alpha=0.7$) and several values of the mass $m_0/m$ and diameter $\sigma_0/\sigma$ ratios. The results are very similar to those found when {a common coefficient of restitution is considered ($\alpha=\alpha_0$).
\begin{figure}
	\begin{center}
	\includegraphics[width=0.98\columnwidth]{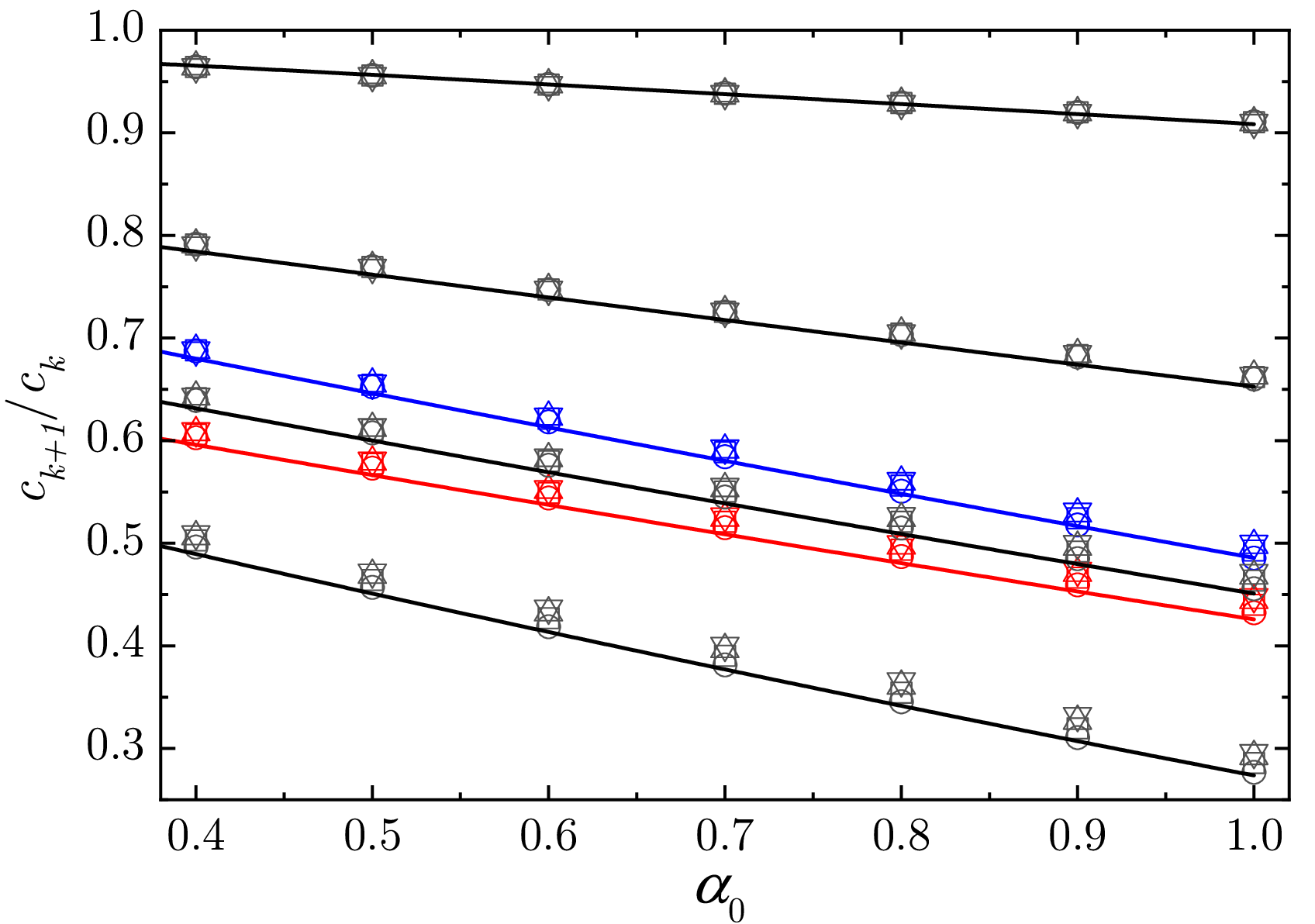}
	\end{center}
	\caption{\vicente{(Color online.)}
		DSMC values of $c_{k+1}/c_k$ vs $\alpha_0$ with $\alpha=0.7$ for $k=1, 2, 3, 4$ (circles, squares, up triangles, down triangles, respectively) and, for from top to bottom,  $\{m_0/m, \sigma_0/\sigma\}=\{6, 1\}$, $\{2, 1\}$  , $\{1, 1/2\}$, $\{1, 1\}$, $\{1, 2\}$ and $\{1/2,1\}$ The solid lines represent $\Omega$ defined by  Eq.~\eqref{Omegafinal}. 
		\vicente{The \{blue, black, red\} color is used for $\sigma_0/\sigma=\{1/2,1,2\}$ }
			\label{fig:ck1ckD}}
	\end{figure}

\subsection{MSD in units of $\la r^2\ra$}

\sby{Respuesta a Reviewer 3, punto 5:}
\vicente{
In this section, we will focus on validating our random walk method by analyzing the MSD in the long-time regime (where $N$ is very large). We will not analyze the MSD for the other time regimes described in Sect.~\ref{sec:ScaRt}.
}

Since the simulated $c_{k+1}/c_k$ values (especially for $k=1$) match the theoretical $\Omega$ well, we expect the theoretical MSD formula \eqref{ele2pa} (which comes from Eq.~\eqref{rbrkok}) to be very  accurate too. This is confirmed in Fig.~\ref{fig:MSDa}  where the ratio $\la R^2 \ra/(\la N \ra \la r^2 \ra)$ is plotted versus the (common) coefficient of restitution $\alpha=\alpha_0$. It is quite apparent that  the MSD per collision  in $\la r^2\ra$ units  is very well described by  the ratio  $1/(1-\Omega)$. 
\begin{figure}
	\begin{center}
	\includegraphics[width=0.98\columnwidth]{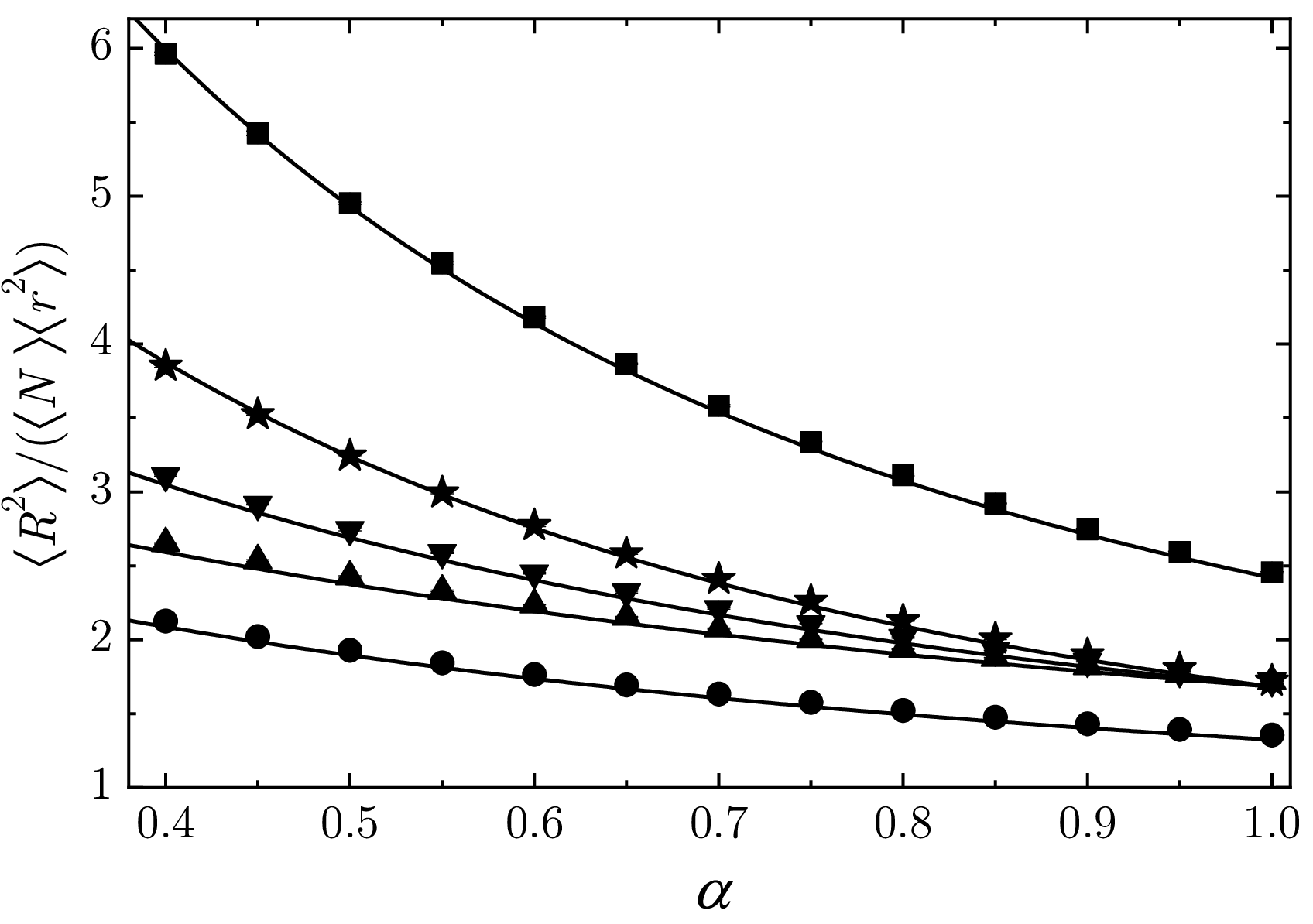}
	\end{center}
	\caption{DSMC values of $\la R^2\ra/(\la N \ra \la r^2\ra)$ vs the (common) coefficient of restitution ($\alpha=\alpha_0$) for, from top to bottom,  
 		$\{m_0/m, \sigma_0/\sigma\}=\{2,1\}$ (squares), $\{1,1/2\}$ (stars), $\{1,1\}$ (down triangles), $\{1,2\}$ (up triangles), and $\{1/2,1\}$ (circles).
		The solid lines represent the function $1/(1-\Omega)$, where $\Omega$ is given by Eq.~\eqref{Omegafinal}.
		\label{fig:MSDa}}
\end{figure}
To complement Fig.~\ref{fig:MSDa}, Fig.~\ref{fig:MSDb} shows  $\la R^2 \ra/(\la N \ra \la r^2 \ra)$ as a function of the coefficient of restitution $\alpha_0$ characterizing the inelasticity of intruder-grains collisions for a given value of $\alpha$ ($\alpha=0.7$).     
Even though we saw good agreement between theory and simulations for the reduced collisional series in Table \ref{tabla1} and Figs.~\ref{fig:ck1ckA}, \ref{fig:ck1ckB}, and \ref{fig:ck1ckD}, the agreement for the MSD between theory and simulation in Figs. \ref{fig:MSDa} and \ref{fig:MSDb} is  very  impressive.
In fact, the agreement is so  surprisingly  good  that one could speculate that the origin of this agreement could be a possible  cancellation or compensation of the approximations leading to Eqs.~\eqref{MSDRW} and \eqref{Omegafinal}. 
Regardless, our MSD results are quite robust, as we have shown in this section and we will see again in Sec.\ \ref{sec:Boltzmann}.
\begin{figure}
	\begin{center}
	\includegraphics[width=0.98\columnwidth]{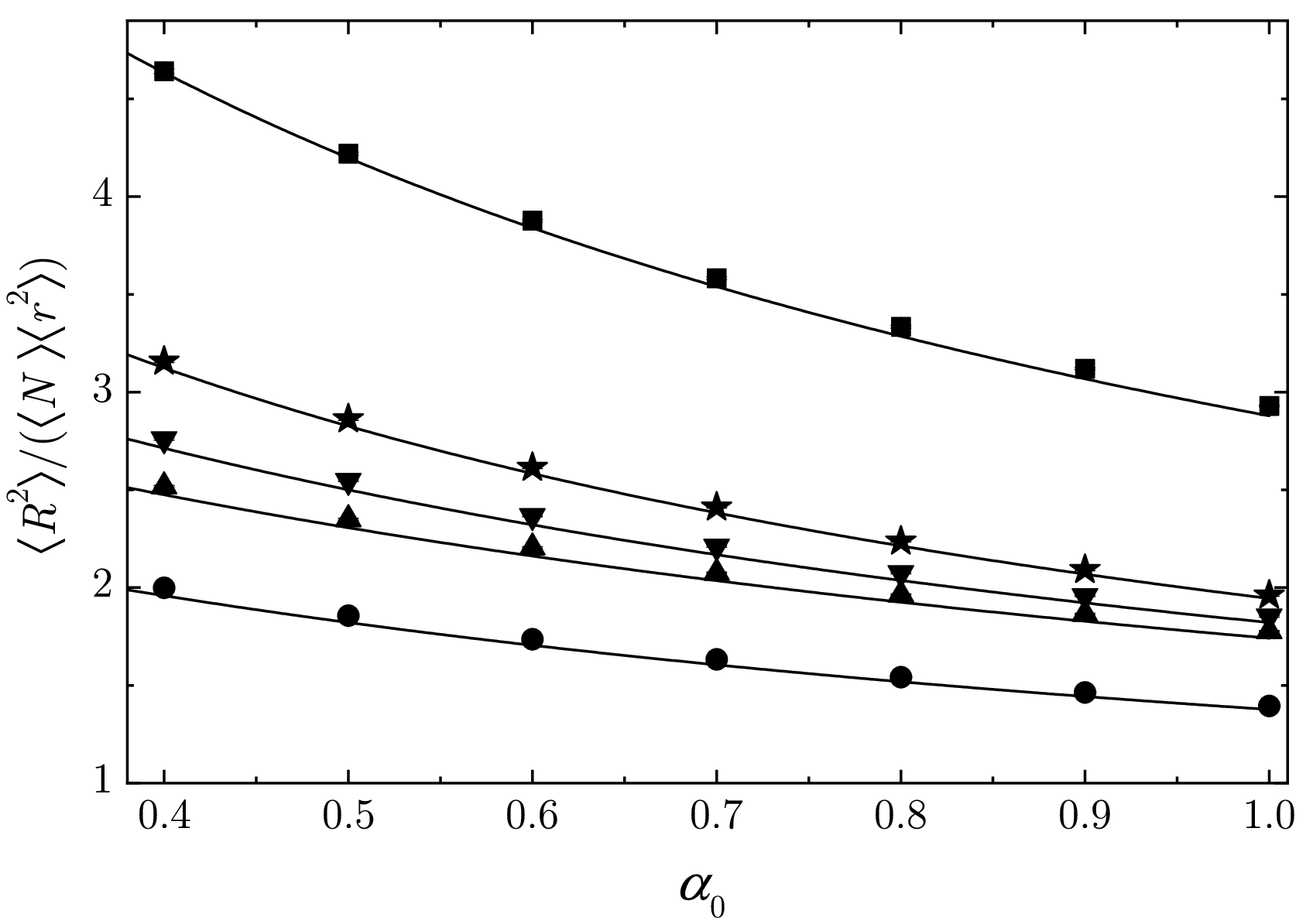}
	\end{center}
	\caption{DSMC values of $\la R^2\ra/(\la N \ra \la r^2\ra)$ vs $\alpha_0$ with $\alpha=0.7$ and  $\{m_0/m, \sigma_0/\sigma\}=\{2,1\}$ (squares), $\{1,1/2\}$ (stars), $\{1,1\}$ (down triangles), $\{1,2\}$ (up triangles), and $\{1/2,1\}$ (circles). The solid lines represent the function $1/(1-\Omega)$, with $\Omega$  given by Eq.~\eqref{Omegafinal}. 		
		\label{fig:MSDb}}
\end{figure}

\subsection{Cancellation of errors}
\sbyt{
As shown in Sec.~\ref{sec:MSDsimu} and as will be seen again in Sec.~\ref{sec:Boltzmann}, our random walk approach leads to remarkably accurate results. In fact, the agreement with simulation results is so surprisingly good that one might reasonably speculate that the origin of this agreement may lie in a cancellation or compensation of errors stemming from the various approximations  leading to Eqs.~\eqref{MSDRW}, \eqref{rbrkok} and \eqref{Omegafinal}.
}

\sbyt{This fortunate cancellation of errors, leading to unexpectedly accurate predictions, is a well-known phenomenon in science in general and in the statistical physics of gases and liquids in particular. To cite just a couple of examples among many: in liquids, the Percus–Yevick approximation outperforms the hypernetted-chain approximation for strongly repulsive and short-ranged potentials, despite retaining fewer diagrams in the diagrammatic expansion \cite{Santos2016book}.  
Another example is found in kinetic theory: replacing the intricate Boltzmann collision operator by the drastic Bhatnagar–Gross–Krook (BGK) approximation often yields remarkably accurate results even for far-from-equilibrium states. For instance, in the case of the uniform shear flow state, the BGK predictions of the shear-rate dependence of the rheological properties of the dilute gas compare quite well with computer simulations \cite{GS03}.   
}

\sbyt{It turns out that our random walk method belongs to the class of theories containing compensating approximations that leads to better-than-expected results. The data presented in Fig.~\ref{fig:v1overv2} justify this claim. This figure displays simulation results of $\langle v_1/v_2\rangle $ and $\langle v_1\rangle/\langle v_2\rangle$, as well as the theoretical approximation to  $\langle v_1/v_2\rangle $ given by Eq.~\eqref{vpvA}. Thus, it serves as a test of   Eq.\eqref{vpvA}.  We see that the percentage error of this  approximation is curiously stable, typically on the order of 13\% (sometimes smaller). 
In any case, this error is clearly larger than that observed in Figs.~\ref{fig:ck1ckA}--\ref{fig:MSDb}. 
Therefore, we may  conclude that the errors arising from the approximations in Secs.~\ref{sec:RW} and \ref{sec:SmoJeans} (which can be identified by the symbol $\approx$ in the equations)---and specifically those in Eqs.~\eqref{r1rkAb},  \eqref{rkfac2}, \eqref{rcprod}, \eqref{rrome}, and \eqref{vpvA}---cancel each other out, leading to final global approximations of our quantities of interest given by Eqs.~\eqref{rbrkok},   \eqref{MSDRWN} [or \eqref{MSDRW}], and \eqref{Omegafinal} that are much better than one might initially expect.
}

\sbyt{ 
Finally, it is worth noting in Fig.~\ref{fig:v1overv2} how accurately our estimate of $\bar{v}_0(t)/\bar{v}_0(t+\tau_0)$, given by the right-hand side of Eq.~\eqref{vpvA}, reproduces the simulation values of $\langle v_1\rangle/\langle v_2\rangle$.
}

\begin{figure}
	\begin{center}
		\includegraphics[width=0.98\columnwidth]{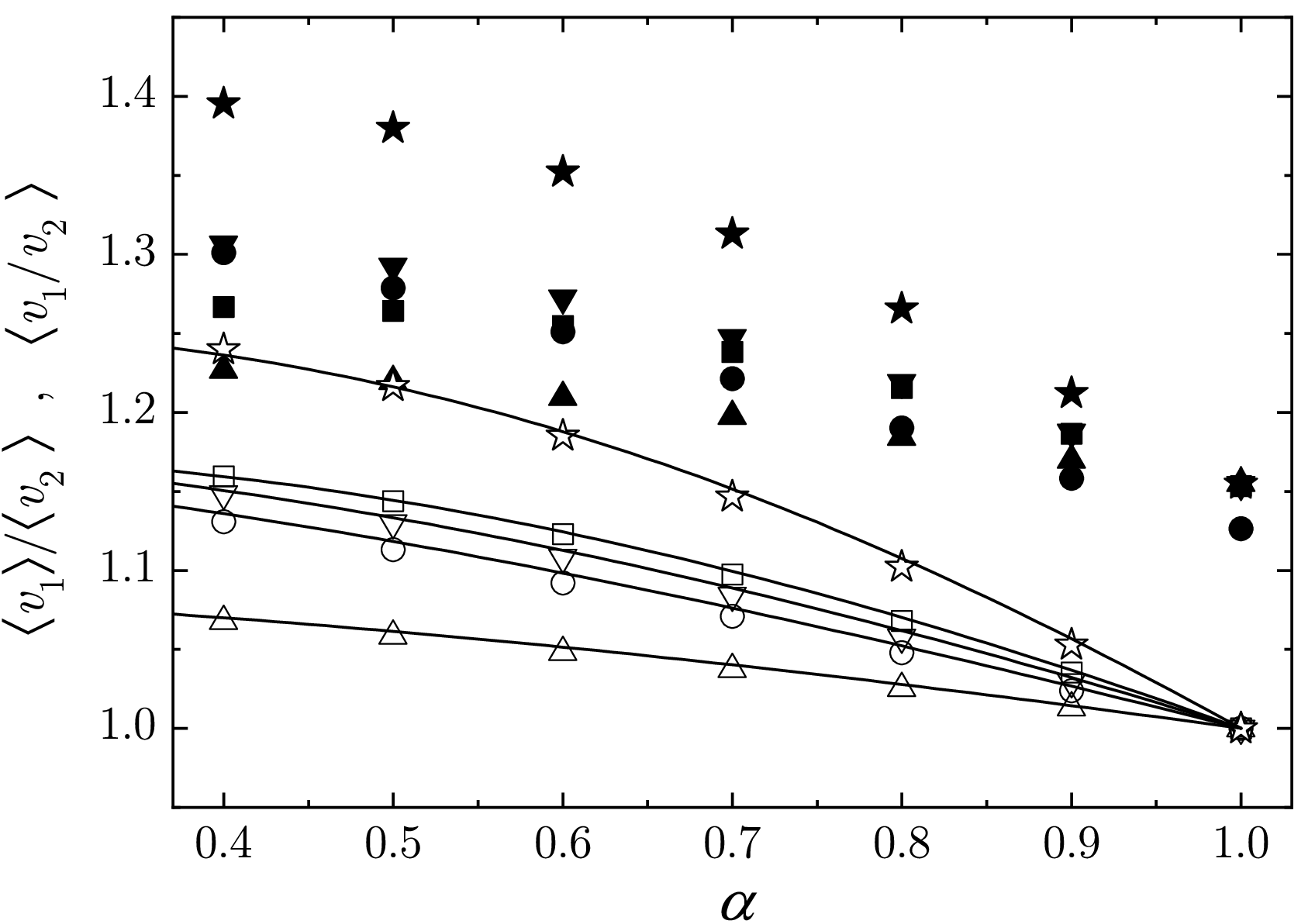}
	\end{center}
	\caption{\sbyt{Simulation results of  $\langle v_1/v_2\rangle $ (filled symbols) and $\langle v_1\rangle/\langle v_2\rangle$ (open symbols)  vs the (common) coefficient of restitution $\alpha=\alpha_0$   for   $\{m_0/m, \sigma_0/\sigma\}=\{2,1\}$ (squares), $\{1,1/2\}$ (stars), $\{1,1\}$ (down triangles), $\{1,2\}$ (up triangles), and $\{1/2,1\}$ (circles). The solid lines represent the theoretical estimation of $\langle v_1\rangle/\langle v_2\rangle$ given by the right-hand side of  Eq.~\eqref{vpvA}.}
		\label{fig:v1overv2}}
\end{figure}

\section{Comparison with the Boltzmann theoretical results}
\label{sec:Boltzmann}

It is natural wonder if the theoretical results derived in this paper for the MSD of an intruder immersed in a granular gas in HCS  (Eqs.~\eqref{ele2pa} and \eqref{Omegafinal}) are comparable, worse, or better than those previously obtained in the granular literature by solving the inelastic version of the Boltzmann kinetic equation by means of the Chapman--Enskog method \cite{CCC70}. As usual in kinetic theory, given that the (time-dependent) tracer diffusion coefficient  $D(t)=\la R^2\ra/2dt$ is given in terms of the solution of a linear integral equation, an explicit expression of $D(t)$ is obtained by considering the first few terms in a Sonine polynomial expansion. If only the leading term in the series expansion is considered, the corresponding expression of $D(t)$ is referred to as the first-Sonine approximation. If one considers two terms in the series expansion, one gets the second-Sonine approximation to the diffusion coefficient. Both Sonine approximations were obtained in Ref.~\cite{GM04} for low-density three-dimensional granular gases and in Ref.~\cite{GV09} for $d$-dimensional granular gases at moderate densities.

 The relationship between the diffusion coefficient $D(t)$ and the effective free path $\ell_e^2=\la R^2\ra/\la N\ra$ is \cite{Abad2022}
\begin{equation}
	\label{ell2lambdaKT}
	\ell_e^2=2d\widetilde{D}\Upsilon\beta \,  \ell_{0,M}^2
	=\frac{2d\widetilde{D}}{\Upsilon} \, \ell_{M}^2,
\end{equation}
where $\widetilde{D}$ is the dimensionless diffusion coefficient defined by
\begin{equation}
	\label{DtildeDef}
	D(t)= 2\left[\frac{\Gamma\left(\frac{d+1}{2}\right)}{\Gamma\left(\frac{d}{2}\right)}\right]^2\, \frac{T(t)}{m\nu(t)}\,  \widetilde{D} .
\end{equation}
Recall that $\ell_{M}\equiv\bar{v}(t)/\nu(t)$ and $\ell_{0,M}\equiv \bar{v}_0(t)/\nu_0(t)$ are the Maxwell mean free paths for the gas particles and intruder, respectively.  Furthermore, note that Eq.~\eqref{ele0M} has been used to get the rightmost equation.  The first- and second-Sonine approximations of the coefficient $\widetilde{D}$ are given in the Appendix \ref{sec:AppB} for the sake of completeness. 

The Maxwell mean free paths may not be the same as the mean free paths $\ell_0$ and $\ell$, which are defined by the average distances traveled by a particle between collisions. However, for elastic hard spheres, $\ell_0=\ell_{0,M}$ and $\ell=\ell_{M}$ \cite{Paik2014}.
Simulations confirm these relations also hold for inelastic hard spheres. Therefore, we will use $\ell_0$ and $\ell$ to denote $\ell_{0,M}$ and $\ell_{M}$, respectively, from now on. Thus, in terms of the dimensionless diffusion coefficient $\widetilde{D}$, the effective free path $\ell_e$ is  \cite{Abad2022}
\begin{equation}
	\label{ell2lambdaKTb}
	\ell_e^2=2d\widetilde{D}\Upsilon\beta \,  \ell_{0}^2
	=\frac{2d\widetilde{D}}{\Upsilon} \, \ell^2,
\end{equation}
where the explicit dependence of $\widetilde{D}$ on the parameter space of the system depends on the specific Sonine approximation considered. 
\sby{ Respuesta Reviewer 2, punto 1.i:}
\vicente{Note that from the relationships $\ell_0=\ell_{0M}$, $\ell=\ell_M$, $\ell_{0M}=\ell_M/(\Upsilon\sqrt{\beta})$ (the latter expression being implicit in Eq.~\eqref{ell2lambdaKT}) combined with the time-independence of  $\ell_M$ (see Eq. \eqref{eleM}), and  the time-independence of  $\Upsilon$ and $\beta$, implies that the free-paths of both the tracer and gas particles are time independent.
} 

\vicente{Equation~\eqref{ell2lambdaKTb} provides the effective free path according to kinetic theory.   
The effective free path}
 $\ell_e$ obtained with our random walk approach is (see Eq.~\eqref{ele2pa})
\begin{equation}
	\label{ele2pb}
	 \ell_e^2=\frac{2\kappa}{1-\Omega}\, \ell_0^2=\frac{2\kappa}{(1-\Omega)\Upsilon^2\, \beta} \,\ell^2.
\end{equation}
Here,
\begin{equation}
	\label{kappaDef}
\kappa\equiv \frac{\la r^2\ra}{ 2 \ell_0^2}.
\end{equation}
The problem we face when comparing the kinetic theory result \eqref{ell2lambdaKTb} with the random walk result \eqref{ele2pb} is that  the relationship between $\la r^2\ra$ and $\la r\ra\equiv \ell_0$ is unknown (that is, $\kappa$ is unknown). For elastic collisions, both the free path distribution $r$  and the ratio $\kappa$ can be expressed as integrals that can be easily evaluated numerically \cite{Yuste2024}. Additionally,} asymptotics expressions for the free path distribution in certain limits are known. In the case of elastic collisions, $\kappa$ depends solely on the mass ratio $m_0/m$ and takes values close to 1 (see Fig.~2 of Ref.~\cite{Yuste2024}). 
For instance,   $\kappa\simeq 1.0506$ for  $m_0/m=1$. 
However, for  granular gases in HCS, no expressions for the free path distribution  and $\kappa$ are known. Therefore,  to offer a clean comparison between  the  kinetic theory 
and the random walk results derived here, 
we resort to  estimate  $\kappa$ from our numerical simulations.

\begin{figure}
	\centering
	\includegraphics[width=0.98\linewidth]{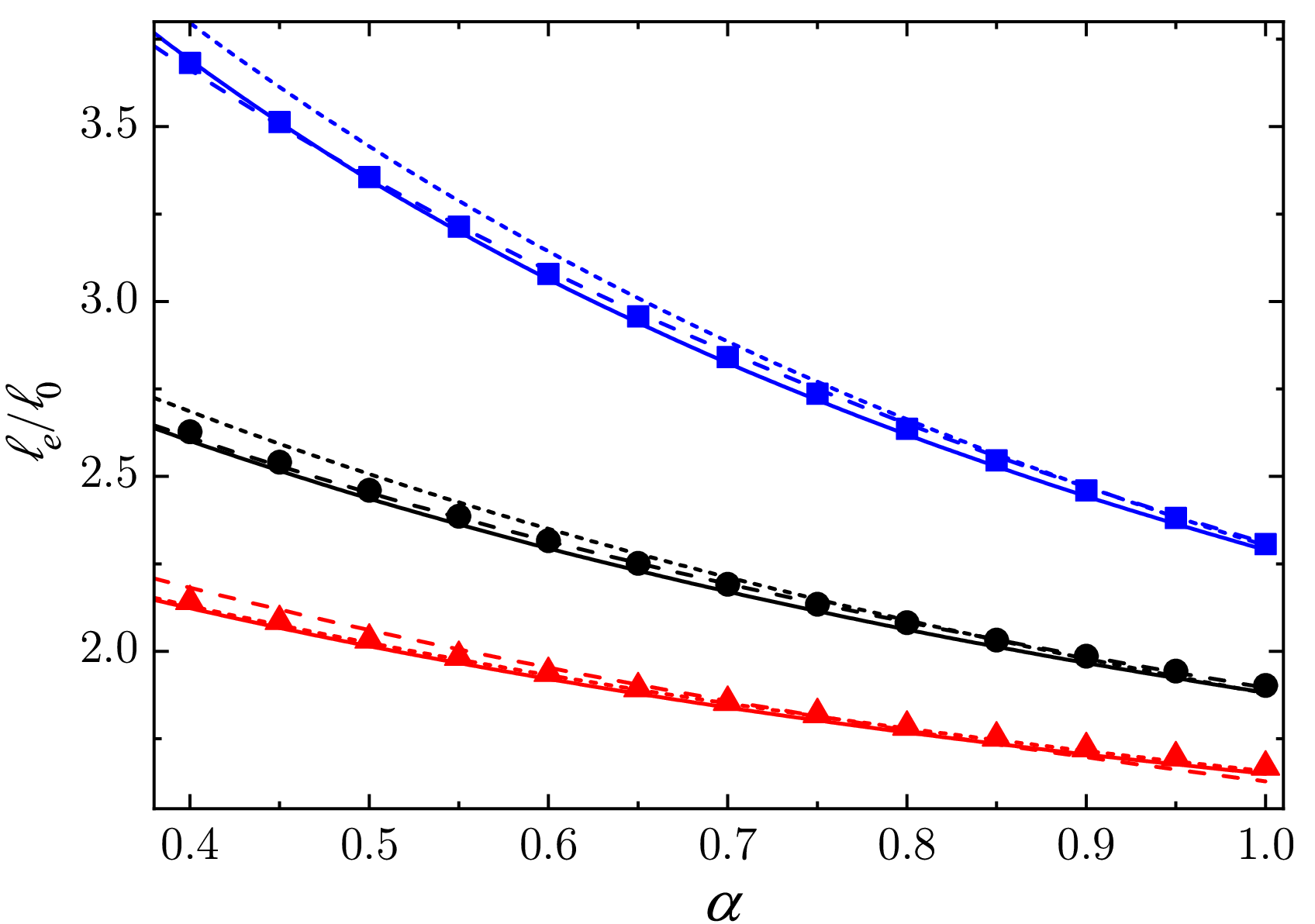}
	\caption{
		(Color online.) 
Ratio of effective mean free path to intruder mean free path, $\ell_e/\ell_0$, as a function of  the (common) coefficient of restitution $\alpha=\alpha_0$ for an intruder  with $\sigma_0/\sigma = 1$ and, from top to bottom, $m_0/m = 2$ \vicente{(blue)}, $m_0/m = 1$ \vicente{(black)}, and $m_0/m = 1/2$ \vicente{(red)}. \vicente{Symbols} represent  DSMC  results,  the solid lines correspond to the random walk results given by Eq.~\eqref{ele2pb}, and the \vicente{short-dashed} and dashed lines refer to the results obtained from the first- and second-Sonine approximations, respectively.   
}
	\label{fig:lem0Var}
\end{figure}

\begin{figure}
\begin{center}
\includegraphics[width=0.98\columnwidth]{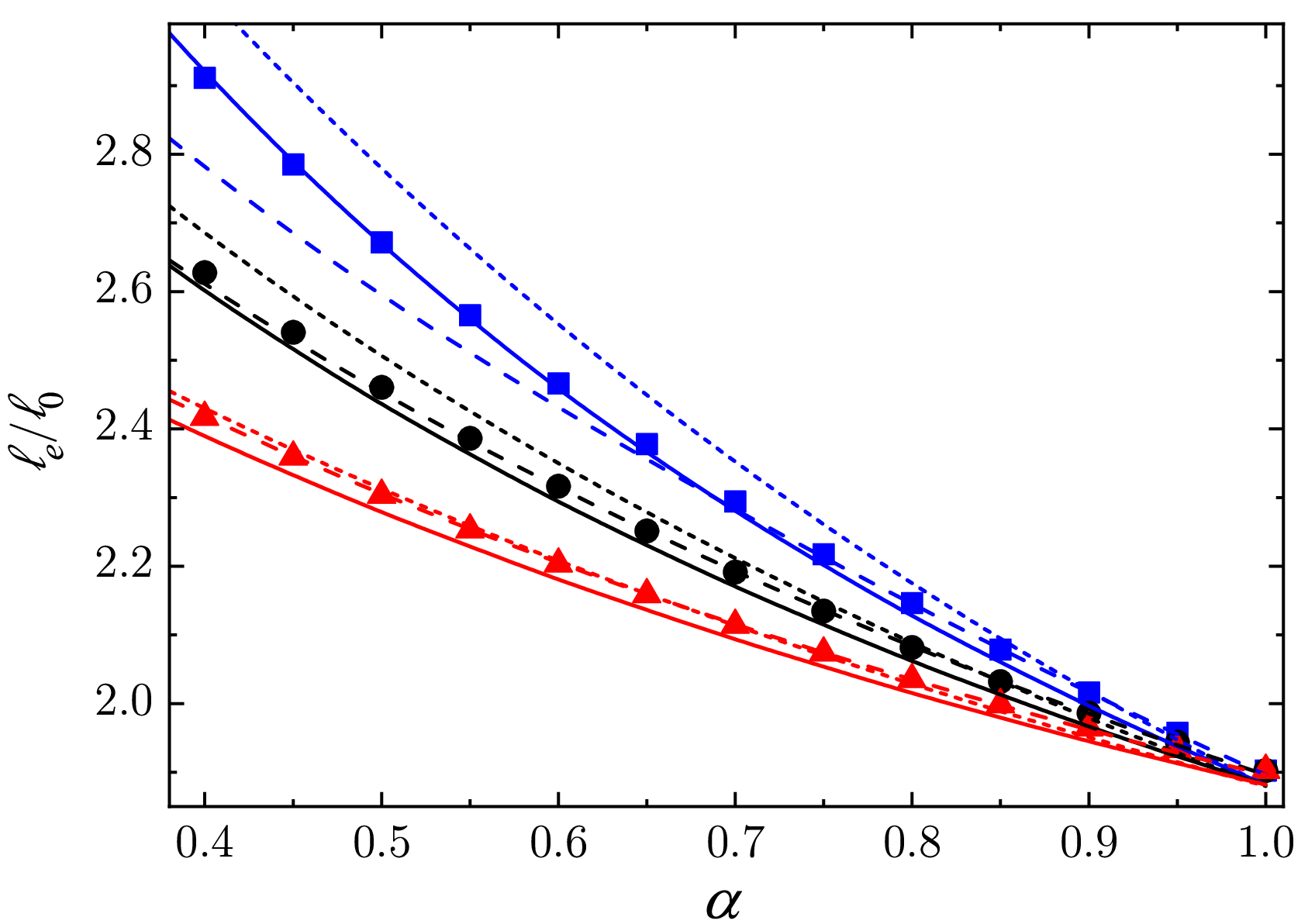}
\end{center}
\caption{(Color online.) 
	Ratio of effective mean free path to intruder mean free path, $\ell_e/\ell_0$, as a function of $\alpha=\alpha_0$ for an intruder  with  $m_0/m = 1$ and, from top to bottom, $\sigma_0/\sigma = 1/2$  \vicente{(blue)}, $\sigma_0/\sigma  = 1$  \vicente{(black)}, and $\sigma_0/\sigma  = 2$  \vicente{(red)}. \vicente{Symbols} represent  DSMC  results,  the solid lines correspond to the random walk results given by Eq.~\eqref{ele2pb}, and the dashed-dotted and dashed lines refer to the results obtained from the first- and second-Sonine approximations, respectively.     
}
\label{fig:lesigma0var}
\end{figure}

Figures \ref{fig:lem0Var} and \ref{fig:lesigma0var} compare simulation results for $\ell_e$ (normalized by the mean free path $\ell_0$) with those obtained from  the theoretical results obtained from  both (i) the first- and second-Sonine  approximations, and (ii) the random walk approximation.
Figure \ref{fig:lem0Var} shows results for  three different mass ratios (intruder mass greater than, equal to, and less than the mass of the granular gas particles) and a common diameter for intruder and gas particles ($\sigma_0=\sigma$). 
We observe that the results from all three approximations are quite good.  While the first-Sonine approximation clearly gives the worst results, the predictions of the second-Sonine approximation and our random walk approximation are very similar and compare quite well with simulations. Figure \ref{fig:lesigma0var} shows the effect of changing the intruder size (intruder diameter less than, equal to, and greater than the gas grain diameter) while keeping the mass ratio fixed ($m_0/m$=1). Here, the agreement of all three approximations with the simulation results is very reasonable, except notably for the case with $\sigma_0/\sigma=1/2$, where we see large differences between the Sonine approximations and the simulation results, especially for high inelasticity.  Surprisingly,  the random walk approximation still performs well  even for quite small values of the (common) coefficient of restitution.
Our results indicate that all approximations tend to improve when the mass  ratio $m_0/m$ is larger than 1 and the diameter ratio $\sigma_0/\sigma$ is smaller than 1.  
This corresponds to cases where  $c_{k+1}/c_k$  takes smaller values, i.e, to cases where the particles become more diffusive, less ballistic, with smaller values of the persistence collisions $\tilde n$.

\section{Conclusions}
\label{sec:Conclu}

This article addresses how far an intruder in a granular gas ({modeled as a gas of inelastic hard spheres) moves when subjected only to collisions with  particles of the granular gas, as the system freely evolves and cools  (cooling due to the inelastic collisions). 
It is known \cite{GM04,GarzoBook19} that the answer to this question can be obtained by solving the (inelastic) Boltzmann kinetic equation with the Chapman-Enskog method. 
Here, we  revisit the problem but employing  a method that can be considered even more classic (older) than the Chapman-Enskog approach, as it originates from the random walk method utilized by Smoluchowski in his analysis of Brownian motion (section 11 of Ref.\ \cite{Smoluchowski1906b}).

The method  followed in this paper is based on noticing that the series for the MSD of a particle after $N$ collisions ($N\gg 1$), 
\beq
\label{7.1}
\la R^2\ra=N\left(\la r^2\ra + 2\la \rb_1\cdot\rb_2\ra+ 2\la \rb_1\cdot\rb_3\ra+\ldots\right)
\eeq
approximates a geometric series with roughly constant ratios between terms. Approximating these ratios by $\Omega$, we obtain the result 
\beq
\label{7.2}
\la R^2\ra= N \la r^2\ra(1+\Omega+\Omega^2+\ldots)=N \frac{\la r^2\ra}{1-\Omega}. 
\eeq
An explicit expression of $\Omega$ (see Eq.~\eqref{Omegafinal}) for a granular gas of   three-dimensional  hard spheres  
has been obtained in this paper. 
This expression  reduces to  the so-called mean persistence ratio  when the collisions are elastic \cite{CCC70}.
To gauge the reliability of this theoretical result, a comparison with the numerical results obtained here from the DSMC method \cite{Bird94} for a granular gas in the HCS has been carried out.
For a wide range of masses, sizes, and coefficients of restitution, the ratio of the successive terms  appearing in Eq.\ \eqref{7.1}} 
is well approximated  (see Table \ref{tabla1} and Figs.~\ref{fig:ck1ckA}, \ref{fig:ck1ckB}, and \ref{fig:ck1ckD})  by our expression \eqref{Omegafinal} for $\Omega$.
This good agreement validates  the accuracy of Eq.\ \eqref{7.2}, 
as shown in Figs.~\ref{fig:MSDa} and \ref{fig:MSDb}. The agreement is surprisingly good.

Finally, we have compared our expression for the MSD with the first- and second-Sonine approximations derived from the Chapman-Enskog solution of the Boltzmann equation. 
The comparison is not straightforward as our MSD expression is formulated in terms of $\la r^2\ra$, while both Sonine approximations are expressed in terms of $\la r \ra$, and the precise relationship between these two quantities is unknown. To enable the comparison, we have used the empirical relationship between $\la r^2\ra$ and $\la r \ra$ obtained from simulations. We found that our MSD expression generally provides significantly better agreement with simulations than the first-Sonine approximation. We found that  in general   our expression  for the MSD compares significantly better with simulations  than the first-Sonine approximation. Moreover, its accuracy is comparable to that of the second-Sonine approximation, which requires the computation of more collision integrals than our random walk approach.

Several extensions of this work are worth exploring. One obvious extension  is to apply our method to hard disks to see how much the dimension of the medium affects its accuracy. Our considerations have been very general, so extending them to other dimensions should be feasible. Specifically, we expect that the mathematical steps  used here to determine the mean persistence ratio  $\la \omega \ra$ in 3D will be quite similar to those required for 2D.  
Furthermore, we have focused on dilute systems with  pair correlations at contact being  1. We could also explore how the results change when granular gases at moderate densities are considered.  
This article have focused on the analysis of the MSD in the long-time regime. An all-time analysis extending beyond the scaling discussion of Sec.~\ref{sec:MSDregimes} would also be of interest.
Also, extending our analysis to the diffusion of an intruder within a granular gas mixture presents an interesting avenue for future research. 
We could also investigate how well our method would describe granular gases with non-hard-core interactions (for instance, by considering the so-called inelastic Maxwell models \cite{GarzoBook19}), or even how the inclusion of rotational degrees of freedom in collisions (inelastic rough hard spheres) would affect the results.  We intend to explore these avenues in the future.

\sby{Texto nuevo motivado por comentario 6 de Reviewer 3:}
\vicente{Finally, it is worth noting that a direct extension of our approach to granular gases subjected to continuous energy injection (e.g., via a thermostat) is not feasible. Our method relies heavily on the concept of \textit{free} paths between collisions and their persistence. However, free paths are not well-defined  (they effectively cease to exist) when  particle trajectories between collisions are perturbed by stochastic interactions originating from an external thermostat.}

\acknowledgments
We acknowledge financial support from Grant No. PID2024-156352NB-I00 funded by MCIU/AEI/10.13039/501100011033/FEDER, UE. 
S.B.Y. and V.G. also acknowledge
financial support from Grant GR24022 funded by Junta de Extremadura (Spain) and by ERDF ``A way of making Europe''. 

\section*{Data Availability}
The data supporting the findings of this study are openly available \cite{datasetHCS}.




\appendix

\section{Evaluation of the mean-persistence ratio}
\label{sec:AppA}

In this Appendix we provide some technical details appearing in the calculation of the mean persistence-ratio of an intruder immersed in a granular gas in HCS. The calculation follows similar steps as those made for molecular gases in the Chapman and Cowling textbook (see chapter 5, sec.\ 5.5 of Ref.\ \cite{CCC70}).

Let us consider an intruder of mass $m_0$ and diameter $\sigma_0$ moving in  a granular gas of mass $m$ and diameter $\sigma$. Let us denote by $\mathbf{v}_2$ and $\mathbf{v}_1$ the post- and pre-collisional velocities of the intruder in an intruder-grain collision. The relationship between $\mathbf{v}_2$ and $\mathbf{v}_1$ is \cite{GarzoBook19}
\beq
\label{a1}
\mathbf{v}_{2}=\mathbf{v}_{1}-\mu\left(1+\alpha_{0}\right)(\widehat{{\boldsymbol {\sigma }}}\cdot {\bf g})\widehat{\boldsymbol {\sigma}},
\eeq
where we recall that $\mu=m/(m+m_0)$ and $\al_{0}$ is the coefficient of normal restitution for intruder-grain collisions. Moreover, $\mathbf{g}=\mathbf{v}_1-\mathbf{v}$ is the relative velocity of the colliding pair ($\mathbf{v}$ is the velocity of the granular gas particle) and $\widehat{{\boldsymbol {\sigma }}}$ is the unit vector along the line joining the centers of the spheres that represent the intruder and the grain at contact. Let $\langle \mathbf{v}_2(\mathbf{v}_1)\rangle$ denote the average (or mean) velocity after collision of the intruder, given that its velocity  before the collision was $\mathbf{v}_1$. 
Taking the average over collisions with particles of the granular gas, one has the result
\beq
\label{a2}
\langle \mathbf{v}_2(\mathbf{v}_1)\rangle=\frac{\overline{\sigma}^{d-1}}{P_{0}} \chi_0 \int d{\bf v}\int d\widehat{\boldsymbol{\sigma}}\,\Theta (\widehat{{\boldsymbol {\sigma}}}\cdot \mathbf{g})(\widehat{\boldsymbol {\sigma }}\cdot
\mathbf{g}) \mathbf{v}_2 f(\mathbf{v}),
\eeq
where $\overline{\sigma}=(\sigma+\sigma_0)/2$, $f(\mathbf{v})$ is the one-particle velocity distribution of the granular gas and \beqa
\label{a3}
P_{0}&=&\overline{\sigma}^{d-1}\chi_0\int d{\bf v}\int d\widehat{\boldsymbol{\sigma}}\,\Theta (\widehat{{\boldsymbol {\sigma}}}\cdot \mathbf{g})(\widehat{\boldsymbol {\sigma }}\cdot
\mathbf{g})f(\mathbf{v})\nonumber\\
&=&B_1 \overline{\sigma}^{d-1}\chi_0\int d{\bf v}\; g\; f(\mathbf{v}).
\eeqa
Here, we have introduced the quantities \cite{NE98}
\beq
\label{a4}
B_k\equiv \int d\widehat{\boldsymbol{\sigma}}\,\Theta (\widehat{{\boldsymbol {\sigma}}}\cdot \widehat{\mathbf{g}})(\widehat{\boldsymbol {\sigma }}\cdot \widehat{\mathbf{g}})^\vicente{k}=\pi^{(d-1)/2}\frac{\Gamma\left(\frac{k+1}{2}\right)}{\Gamma\left(\frac{k+d}{2}\right)}.
\eeq
Equation \eqref{a2} can be more explicitly written when one takes into account the scattering rule \eqref{a1}:
\beq
\label{a5}
\langle \mathbf{v}_2(\mathbf{v}_1)\rangle=\mathbf{v}_1-B_3 \frac{\overline{\sigma}^{d-1}}{P_{0}}\chi_0\mu(1+\al_{0})\int d\mathbf{v}\; g\; \mathbf{g}\; f(\mathbf{v}),
\eeq
where use has been made of the result
\beq
\label{a6}
\int d\widehat{\boldsymbol{\sigma}}\,\Theta (\widehat{{\boldsymbol {\sigma}}}\cdot \mathbf{g})(\widehat{\boldsymbol {\sigma }}\cdot \mathbf{g})^k \widehat{{\boldsymbol {\sigma}}}=B_{k+1}g^{k-1}\mathbf{g}.
\eeq
Since $\mathbf{g}=\mathbf{v}_1-\mathbf{v}$, then Eq.\ \eqref{a2} becomes
\beqa
\label{a7}
\langle \mathbf{v}_2(\mathbf{v}_1)\rangle&=&\Big[1-\frac{2}{d+1}\mu(1+\al_{0})\Big]\mathbf{v}_1+\frac{2B_1}{d+1}
\frac{\overline{\sigma}^{d-1}}{P_{0}}\nonumber\\
& & \times\chi_0\mu(1+\al_{0})\int d\mathbf{v} g\; \mathbf{v}\; f(\mathbf{v}),
\eeqa
where the identity $B_3=(2/(d+1))B_1$ has been used.

For the sake of concreteness, henceforth we consider a three-dimensional ($d=3$) system. In this case, to evaluate the integral appearing in Eq.\ \eqref{a7}, we express $\mathbf{v}$ in terms of spherical coordinates $v$, $\theta$, and $\varphi$ about $\mathbf{v}_1$ as axis. Since the granular gas is in the HCS, its velocity distribution $f(\mathbf{v})$ depends on $\mathbf{v}$ only through its modulus $v$. This means that the only nonvanishing contribution in the integral appearing in \eqref{a7} comes from the component $v \cos \theta$ of $\mathbf{v}$ in the direction of $\mathbf{v}_1$.
 Thus, from Eq.\ \eqref{a7}, the mean value of $\mathbf{v}_2$ can be written as
\beq
\label{a8}
\langle \mathbf{v}_2(\mathbf{v}_1)\rangle=\omega \mathbf{v}_1,
\eeq
where $\omega$   is 
\begin{widetext}
\beqa
\label{a9}
\omega&=&1-\frac{1}{2}\mu(1+\al_{0})+\frac{\pi}{2}\frac{\overline{\sigma}^{2}}{P_{0}}\chi_0 \mu(1+\al_{0})\int d\mathbf{v}\; g \;\frac{v\cos \theta}{v_1} f(\mathbf{v})\nonumber\\
&=&
1-\frac{1}{2}\mu(1+\al_{0})+\pi^2\frac{\overline{\sigma}^{2}}{P_{0}}\chi_0\mu(1+\al_{0})
\int_0^{\infty}dv \frac{v^3}{v_1}f(v)\int_0^\pi g\;  \sin \theta \cos \theta\; d\theta .
\eeqa
\end{widetext}
The quantity $\omega(v_1)$ gives the ratio of the mean value of the velocity of a particle after collision to the velocity before collision when the latter velocity is of magnitude $v_1$. It can be referred to as the persistence-ratio for particles of speed $v_1$ \cite{CCC70}. To perform the angular integral in Eq.\ \eqref{a9}, one has take into account the identity
\begin{widetext}
\beq
\label{a10}
g=\sqrt{v^2 \sin^2 \theta \cos^2 \varphi+v^2 \sin^2 \theta \sin^2 \varphi+(v_1-v \cos \theta)^2}=v \sqrt{\sin^2\theta+\left(\frac{v_1}{v} -\cos \theta\right)^2}.
\eeq
Using \eqref{a10}, the angular integral in Eq.\ \eqref{a9} gives
\beq
\label{a11}
\int_0^\pi d\theta \sin \theta \cos \theta\sqrt{\sin^2\theta+\left(A -\cos \theta\right)^2}=\left\{
\begin{array}{cc}
\frac{2}{15}A\left(A^2-5\right),&\text{if } A<1,\\
\frac{2}{15}\frac{1-5A^2}{A^2},& \text{if } A>1,
\end{array}
\right.
\eeq
where $A\equiv v_1/v$. Taking into account the above results, $\omega$ can be written as
\beq
\label{a12}
\omega(v_1)=1-\frac{1}{2}\mu(1+\al_{0})+\frac{2}{15}\pi^2 \frac{\overline{\sigma}^{2}}{P_{0}}\chi_0\mu(1+\al_{0})\left\{\int_0^{v_1}dv\frac{v^4}{v_1^3}\left(v^2-5v_1^2\right)f(v)+
\int_{v_1}^\infty dv v\left(v_1^2-5v^2\right)f(v)\right\}.
\eeq
\end{widetext}

We are now in conditions to determine the average of  $\omega(v_1)$  over all possible values of $v_1$.
 This is nothing else than the mean-persistence ratio $\la \omega \ra$ defined as \cite{CCC70}
\beq
\label{a13}
\la \omega \ra=\frac{\int d\mathbf{v}_1\;\omega(v_1)\; P_{0}(\mathbf{v}_1)\; f_0(\mathbf{v}_1)}{N_0},
\eeq
where $f_0(\mathbf{v}_1)$ is the one-particle velocity distribution of intruders and $N_0$ is the total number of collisions between the intruder and granular gas particles per unit time. This quantity is defined as
\beq
\label{a14}
N_{0}=\int d\mathbf{v}_1\; P_{0}(\mathbf{v}_1)f_0(\mathbf{v}_1).
\eeq
Hence, the mean-persistence $\la \omega \ra$ is given by
\begin{widetext}
\beqa
\label{a15}
\langle \omega \rangle&=&
1-\frac{1}{2}\mu(1+\al_{0})+\frac{2}{15}\pi^2 \frac{\overline{\sigma}^{2}\chi_0\mu(1+\al_{0})}{N_{0}}\int d\mathbf{v}_1\; f_0(\mathbf{v}_1)\Bigg[\int_0^{v_1}dv\frac{v^4}{v_1^3}\left(v^2-5v_1^2\right)f(v)+\int_{v_1}^\infty dv v\left(v_1^2-5v^2\right)f(v)\Bigg]\nonumber\\
&=&1-\frac{1}{2}\mu(1+\al_{0})+\frac{8}{15}\pi^3 \frac{\overline{\sigma}^{2}\chi_0\mu(1+\al_{0})}{N_{0}}\int_0^\infty dv_1\; v_1^2 f_0(v_1)
\Bigg[\int_0^{v_1}dv\frac{v^4}{v_1^3}\left(v^2-5v_1^2\right)f(v)\nonumber\\
& &+
\int_{v_1}^\infty dv v\left(v_1^2-5v^2\right)f(v)\Bigg],
\eeqa
\end{widetext}
where in the last step we have accounted for that $f$  
is isotropic in $\mathbf{v}_1$ in the HCS. So far, the expression \eqref{a15} for $\la \omega \ra$ is exact for inelastic hard spheres. However, to estimate it one approximates $f$ and $f_0$ by the Maxwellian distributions
\beq
\label{a16}
f(\mathbf{v})\to f_\text{M}(\mathbf{v})=n \left(\frac{m}{2\pi T}\right)^{3/2} \exp \left(-\frac{m v^2}{2T}\right),
\eeq  
\beq
\label{a17}
f_0(\mathbf{v})\to f_\text{0,M}(\mathbf{v})=n_0 \left(\frac{m_0}{2\pi T_0}\right)^{3/2} \exp \left(-\frac{m_0 v^2}{2T_0}\right),
\eeq 
where $n$ and $n_0$ are the number densities of intruders and granular gas particles. 
For simplicity,   we have set  Boltzmann's constant to 1,  $k_\text{B}=1$, in Eqs.~\eqref{a16} and \eqref{a17}.
 The substitution of these Maxwellian distributions into Eq.\ \eqref{a14} allows us to get $N_0$. It is given by   \cite{Abad2022} 
  \beq
 \label{a18}
 N_{0}=n_0\nu_0,
 \eeq
 where $\nu_0$ is given by Eq.~\eqref{nu0nu}.
To perform the integrals in \eqref{a15}, we introduce the dimensionless velocities $\mathbf{c}=\mathbf{v}/v_\text{th}$ and $\mathbf{c}_1=\mathbf{v}_1/v_\text{th}$. In terms of $c=\eta c_1$, the mean-persistence ratio $\la \omega \ra$ is 
\begin{widetext}
\beq
\label{a19}
\la \omega \ra=1-\frac{1}{2}\mu(1+\al_{0})+\frac{8}{15}\frac{\overline{\sigma}^{2}\chi_0\mu(1+\al_{0})}{N_{0}}
n n_0 v_\text{th} \beta^{3/2}\Big[\int_0^1 d\eta \eta^4\left(\eta^2-5\right)I(\eta)+\int_1^\infty d\eta \eta\left(1-5\eta^2\right)I(\eta)\Big],
\eeq
where
\beq
\label{a20}
I(\eta)=\int_0^\infty\; dx\; x^6\; e^{-(\eta^2+\beta)x^2}=\frac{15}{16}\frac{\sqrt{\pi}}{\left(\eta^2+\beta\right)^{7/2}}.
\eeq
Thus, $\la \omega \ra$ can be written as
\beq
\label{a21}
\la \omega \ra=1-\frac{1}{2}\mu(1+\al_{0})+\frac{1}{2}\sqrt{\pi}\frac{\overline{\sigma}^{2}\chi_0\mu(1+\al_{0})}{N_{0}}
n n_0 v_\text{th} \beta^{3/2}\Bigg[\int_0^1 d\eta \frac{\eta^4\left(\eta^2-5\right)}{\left(\eta^2+\beta\right)^{7/2}}+\int_1^\infty d\eta \frac{\eta\left(1-5\eta^2\right)}{\left(\eta^2+\beta\right)^{7/2}}\Bigg].
\eeq
The final expression of the mean-persistence ratio can be achieved when one evaluates the integrals and inserts Eq.\ \eqref{a18} into Eq.\ \eqref{a21}. The result is
\beq
\label{a22}
\la \omega \ra=1-\frac{1}{2}\mu(1+\al_{0})+\frac{1}{4}\mu(1+\alpha_{0})\beta^{3/2}\left(\frac{\beta}{1+\beta}\right)^{1/2}
\frac{\beta\sqrt{1+\beta}\sinh^{-1}\left(\beta^{-1/2}\right)-1-\beta}{\beta\sqrt{1+\beta}}.
\eeq
\end{widetext}
It is straightforward to see that Eq.\ \eqref{omegral} is  equivalent to expression \eqref{a22}. For elastic collisions ($\alpha=\alpha_{0}=1$), $\beta=m_0/m$ and Eq.\ \eqref{a22} (or its equivalent Eq.\ \eqref{omegral}) yields
\beq
\label{a23}
\la \omega \ra=\frac{1}{2}\mu_\vicente{0}+\frac{1}{2}\mu_{0}^2\mu^{-1/2}\ln \left(\frac{1+\mu^{1/2}}{\mu_{0}^{1/2}}\right),
\eeq
where $\mu_0=m_0/(m+m_0)$. Equation \eqref{a23} agrees with the result (5.51,1) obtained in the Chapman--Cowling textbook \cite{CCC70} for molecular gases. In particular, when $m_0=m$, Eq.\ \eqref{a23} yields the simple result $\la \omega \ra \simeq 0.405806$. When the intruder and granular gas particles are mechanically equivalent ($m_0/m=\sigma_0/\sigma=1$ but $\alpha=\alpha_{0}\neq 1$), $\beta=1$ and Eq.\ \eqref{a22} leads to
\beqa
\label{a24}
\la \omega \ra&=&\frac{3-\al}{4}+\frac{1+\al}{16}\left(\sqrt{2}\sinh^{-1}(1)-2\right)\nonumber\\
&\simeq& 1-0.2971(1+\al).
\eeqa
In this case $\Omega$ takes this simple form:
\begin{equation}
    \Omega\simeq \frac{13-\alpha^2}{11+\alpha^2}\left[1-0.2971(1+\al)\right].
\end{equation}

\vicente{
\section{Alternative expressions for the velocity ratio}
\label{App:IntRules}

\sby{Sección nueva como respuesta a Reviewer 2 punto 4:}

This appendix considers two alternative expressions to Eq.~\eqref{v0e2} for the velocity ratio $\bar{v}_0(t)/\bar{v}_0(t+\tau_0)$. We then show how these changes would affect the theoretical estimates of $c_{k+1}/c_k$ and the  MSD.  

Equation \eqref{v0e2} was derived from Equation \eqref{v0e1} by approximating the integral using the trapezoidal rule. A simpler alternative is to approximate this integral with the rectangle rule. If we use the left rectangle rule, Eq.~\eqref{v0e2} would take the form $\bar{v}_0(t+\tau_0)-\bar{v}_0(t)\approx\zeta(t)\tau_0(t) \bar{v}_0(t)/2$, from which it follows that
\begin{equation}
	\label{v0e2LRR}
	\frac{\bar{v}_0(t)}{\bar{v}_0(t+\tau_0)}\approx\left[1-\zeta \tau_0/2\right]^{-1}.
\end{equation}
Inserting Eq.~\eqref{zetanu0} into this equation and taking into account Eqs.~\eqref{OmegaDef} and \eqref{v1v2aprox}, we find 
\begin{equation}
	\label{xgralEB}
\Omega_\text{LRR}=	 \frac{\langle \omega \rangle}{1-(1-\alpha^2)/2d\Upsilon},
\end{equation}
where the subscript LRR means the left rectangle rule.

Similarly, if we use the right rectangle rule in Eq.~\eqref{v0e2}, we get $\bar{v}_0(t+\tau_0)-\bar{v}_0(t)\approx\zeta(t+\tau_0)\tau_0(t) \bar{v}_0(t+\tau_0)/2$, which implies
\begin{equation}
	\label{v0e2RRR}
	\frac{\bar{v}_0(t)}{\bar{v}_0(t+\tau_0)}\approx 1+\zeta \tau_0/2.
\end{equation}
Again, by inserting Eq.~\eqref{zetanu0} into this equation and considering Eqs.~ \eqref{OmegaDef} and \eqref{v1v2aprox}, we obtain the following relationship:
\begin{equation}
	\label{xgralEC}
	\Omega_\text{RRR}=	  \left( 1+\frac{1-\alpha^2}{2d\Upsilon}\right) \langle \omega \rangle,
\end{equation}
where here the subscript RRR means the right rectangle rule.

In Fig.~\ref{fig:ck1ckABC}, we compare $\Omega_\text{LRR}$ and $\Omega_\text{RRR}$ with the perseverance $\Omega$ given by Eq.~\eqref{Omegafinal} and with simulation data for $c_{k+1}/c_k$ for three cases we considered in Fig.~\ref{fig:ck1ckA}.  We see that the differences among the three theoretical expressions are very small.   
\begin{figure}
\begin{center}
	\includegraphics[width=0.98\columnwidth]{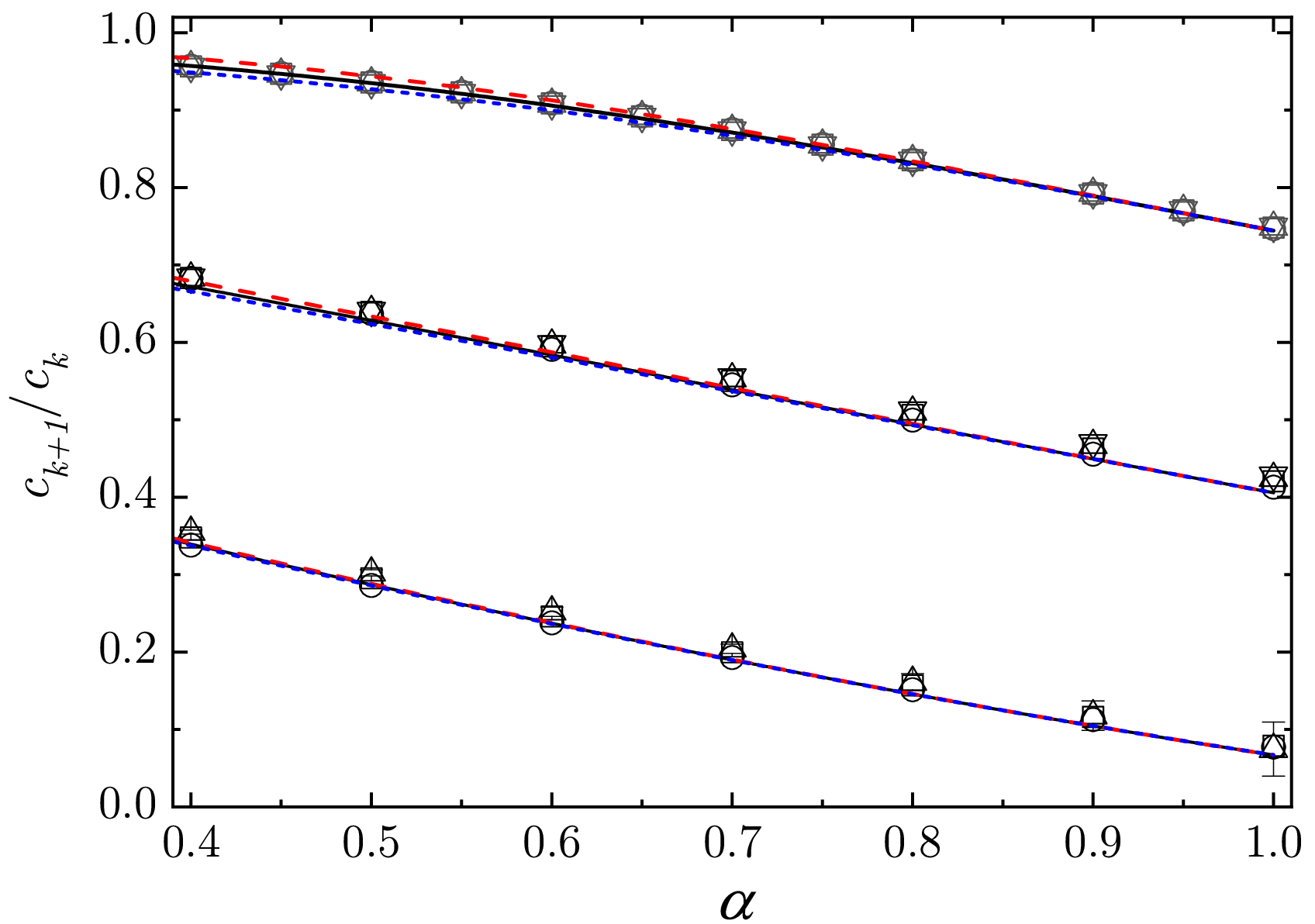}%
\end{center}
\caption{
	\vicente{ (Color online.) 
DSMC values of the first ratios $c_{k+1}/c_k$ of the reduced collisional series vs the (common) coefficient of restitution $\alpha=\alpha_0$ for $k=1, 2, 3, 4$ (circles, squares, up triangles, down triangles, respectively) and, for from top to bottom, $m_0/m= 4, 1,  1/8$ with $\sigma_0/\sigma=1$ (see Fig.\ref{fig:ck1ckA} for more details). 	The \vicente{black} solid lines represent $\Omega$ defined by  Eq.~\eqref{Omegafinal}, the \vicente{red} dashed lines are $\Omega_\text{LRR}$ and the \vicente{blue} short-dashed lines are $\Omega_\text{RRR}$.
}
		\label{fig:ck1ckABC}}
\end{figure}

In Fig.~\ref{fig:MSDABC} we compare the scaled MSD simulation values with theoretical predictions derived from the persistence expressions  $\Omega_\text{LRR}$,  $\Omega_\text{RRR}$ \vicente{and} $\Omega$ (the latter given by Eq.~\eqref{Omegafinal}).  Again, the differences are small, especially when the restitution coefficients  are close to one. We also observe that for the case with a larger intruder mass, the expression from Eq.~\eqref{Omegafinal} (obtained via the trapezoidal rule) is significantly better than those derived using the rectangular rules.

\begin{figure}
	\begin{center}
		\includegraphics[width=0.98\columnwidth]{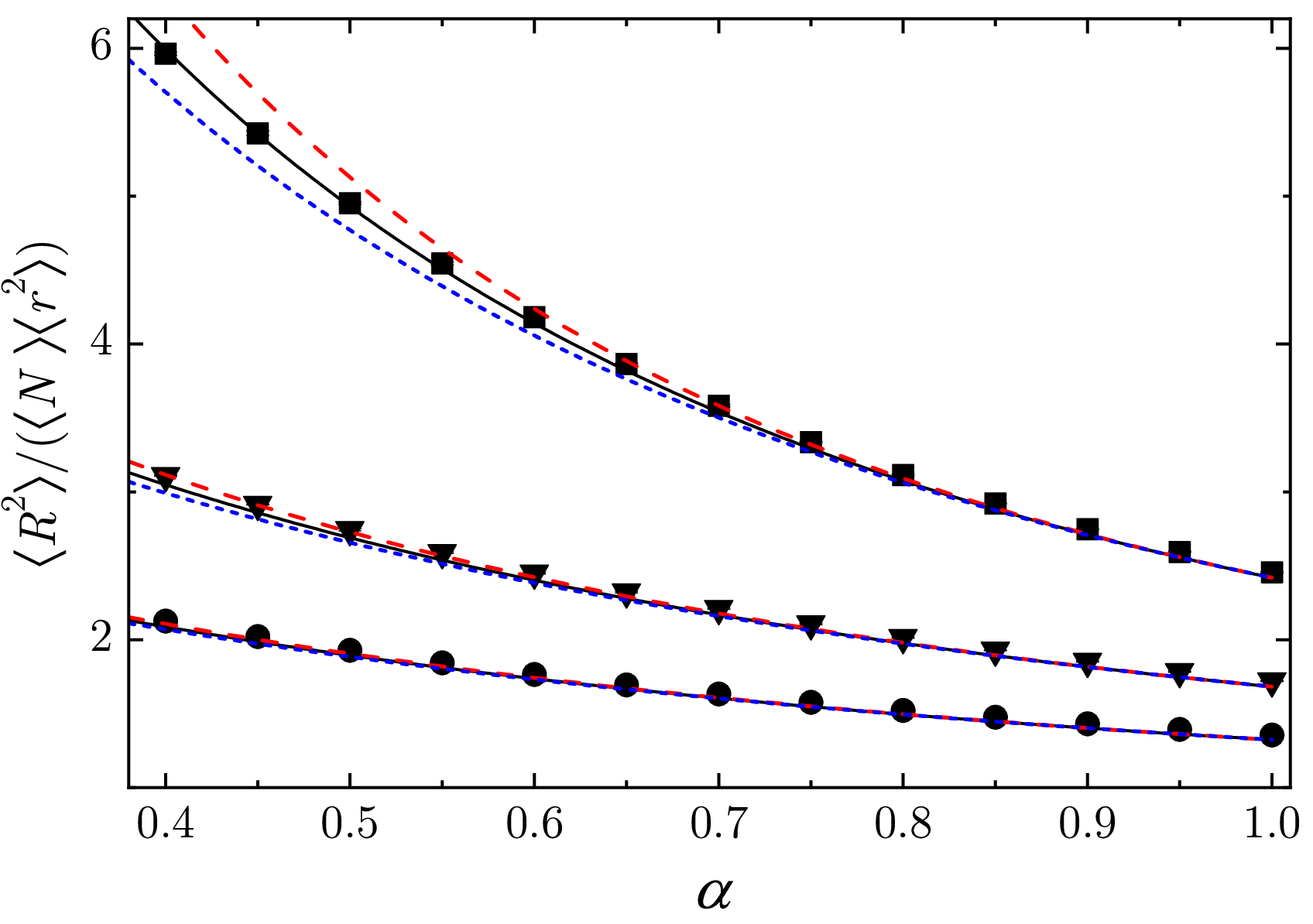}%
	\end{center}
	\caption{
\vicente{ (Color online.) 
				DSMC values of $\la R^2\ra/(\la N \ra \la r^2\ra)$ vs the (common) coefficient of restitution ($\alpha=\alpha_0$) for, from top to bottom,  
		$\{m_0/m, \sigma_0/\sigma\}=\{2,1\}$ (squares),  $\{1,1\}$ (down triangles), and $\{1/2,1\}$ (circles).
		The   lines represent the function $1/(1-\Omega)$, where $\Omega$ is given by Eq.~\eqref{Omegafinal} (black solid line), $\Omega_\text{LRR}$ (red dashed line) and $\Omega_\text{RRR}$ (blue short-dashed line).
	}
		\label{fig:MSDABC}}
\end{figure}
}

\section{Direct simulation Monte Carlo method}
\label{sec:AppB} 

The details of the DSMC method are well-documented in Ref.~\cite{Yuste2024} for a molecular gas in equilibrium. In the HCS, however, grains and intruders lose energy due to the inelasticity of collisions. Therefore, it is necessary to account for this energy loss in the velocity changes after collisions and in the decay of temperature. However, if the granular mixture (in our case, tracers and granular gas) is excited by the Gaussian thermostat
\beq\label{therm}
F_i = -\frac{1}{2} m_i \mathbf{v}\zeta_i
\eeq
which exactly compensates for the collisional energy loss, the measured quantities remain the same as for the HCS, and the temperature achieve a stationary value which makes the DSMC implementation simpler. \vicente{The value of $\zeta_i$ corresponds to that obtained from kinetic theory (see, for example, Ref.~\cite{GarzoBook19}). This coefficient is kept constant throughout the simulation, effectively modeling a steady cooling rate. As a result, the instantaneous kinetic temperature of the intruder, defined as
$$
T_0(t) = \frac{m_0}{d} \left\langle v_i^2(t) \right\rangle_i,
$$
where the average $\langle \cdot \rangle_i$ is taken over all intruder particles at time $t$, eventually reaches a steady value.}
Apart from this, step 4 detailed in Ref.~\cite{Yuste2024} must be slightly modified to incorporate the restitution coefficient. Hence, if the collision is accepted, the velocities of particles are updated according to the scattering rules \cite{GarzoBook19}:
    \beqa
& &\mathbf{v}_k\to\mathbf{v}_k-(1+\alpha_{ij})\mu_{ji}({\mathbf g}_{k\ell}\cdot\widehat{\boldsymbol{\sigma}}_{k\ell})\widehat{\boldsymbol{\sigma}}_{k\ell},\nonumber\\
& &\mathbf{v}_\ell\to\mathbf{v}_\ell+(1+\alpha_{ij})\mu_{ij}({\mathbf g}_{k\ell}\cdot\widehat{\boldsymbol{\sigma}}_{k\ell})\widehat{\boldsymbol{\sigma}}_{k\ell},
\eeqa
being $\mu_{ij}=m_i/(m_i+m_j)$,  $\widehat{{\boldsymbol {\sigma }}}_{k\ell}$ a random colliding direction for a pair of colliding particles labeled as $k$ and $\ell$, and ${\mathbf g}_{k\ell}=\mathbf{v}_k-\mathbf{v}_\ell$. Here, $i$ and $j$ label the particles of species $i$ and $j$ that are selected as candidates to collide in each Monte Carlo step.

In this specific work, \vicente{we consider $10^5$ intruder particles. These particles do not interact significantly with the bulk granular gas, as their concentration is negligible. Their role is purely statistical: by simulating a large number of intruders, we obtain accurate averages of variables related to their dynamics (e.g., velocity distribution, temperature). The total number of simulated particles is approximately $3 \times 10^5$, but only $\sim 2 \times 10^5$ of them correspond to the excess granular particles. At each DSMC time step, we propose, on average, 400 candidate collision pairs. This number is chosen to ensure that the number of accepted collisions statistically reproduces the expected Boltzmann collision rate. The process is repeated up to $3 \times 10^5$ runs to achieve good statistical accuracy}.

\sby{Texto respuesta a punto 2.iii  Reviewer 2}
\vicente{
In order to estimate the standard errors of Table I we have proceeded as follows. We ran  $M$ times the program  described in the previous paragraph and we record the $M$ values of $\langle \rb_{1}\cdot \rb_{k}\rangle$ for $k=1,\ldots, 5$ from each run.  For each $k$, we then calculated the mean $a_k$ of $\langle \rb_{1}\cdot \rb_{k}\rangle$  and the standard deviation, $s_k$, of these  $\langle \rb_{1}\cdot \rb_{k}\rangle$  values over these $M$ runs. Next, we estimated the ratio $c_{k+1}/c_{k}$ from the values of $a_k$.  
The standard deviation of  $c_{k+1}/c_{k}$ was derived from $s_k$ using standard error propagation formulas for a division.  In our simulations, the values of $M$ are of order of a few tens.  For example, for the cases from Table I where   $m_0/m=1$,   $\sigma_0/\sigma=1$, and $\alpha=\alpha_0=0.4,0.6,0.8,1$, the corresponding values of $M$ were 18, 17, 22, 20, respectively.  In our simulations we never used an $M$ value smaller than 10.
}

\section{First- and second-Sonine approximations to the diffusion coefficient}
\label{sec:Appc}

In this Appendix we display the explicit expressions of the (dimensionless) diffusion coefficient $\widetilde{D}$ derived by considering the first- and second-Sonine approximations. 

The first-Sonine approximation $\widetilde{D}[1]$ to the coefficient $\widetilde{D}$ is given by \cite{GM04,GV09,Abad2022}
\beq
\label{c1}
\widetilde{D}[1]=\frac{1}{2}\Bigg[\frac{\Gamma\left(\frac{d}{2}\right)}{\Gamma\left(\frac{d+1}{2}\right)}\Bigg]^2 \frac{T_0/T}{(\nu_a^*-\frac{1}{2}\zeta^*)},
\eeq
where
\beq
\label{c2}
\zeta^*=\frac{\zeta}{\nu}=\frac{1-\alpha^2}{d}
\eeq
is the (dimensionless) cooling rate and $\nu_a^*=\nu_a/\nu$ is the dimensionless collision frequency \cite{Abad2022}
\beq
\label{c3}
\nu_a^*=\frac{\sqrt{2}}{d} \left(\frac{\bar{\sigma}}{\sigma}\right)^{d-1}\frac{\chi_0}{\chi} \mu (1+\alpha_0) \left(\frac{1+\beta}{\beta}\right)^{1/2}.
\eeq

The first-Sonine approximation $\widetilde{D}[2]$ to the coefficient $\widetilde{D}$ can be written as \cite{GM04,GV09}
\beq
\label{c4}
\widetilde{D}[2]=\Lambda \widetilde{D}[1],
\eeq
where $\Lambda$ is 
\beq
\label{c5}
\Lambda=\Bigg[1+\frac{\nu_b^*(\zeta^*-\nu_c^*)}{(\nu_a^*-\frac{1}{2}\zeta^*)(\nu_d^*-\frac{3}{2}\zeta^*)}\Bigg]
^{-1}.
\eeq
Here, we have introduced the (dimensionless) collision frequencies
\begin{widetext}
\begin{equation}
\label{c6}
\nu_{b}^*=\frac{1}
{\sqrt{2}d}\left(\frac{\overline{\sigma}}{\sigma}\right)^{d-1}\frac{\chi_0}{\chi}\mu(1+\alpha_0)[\beta(1+\beta)]^{-1/2},
\end{equation}
\begin{equation}
\label{c7}
\nu_{c}^*=\frac{\sqrt{2}}
{d(d+2)}\left(\frac{\overline{\sigma}}{\sigma}\right)^{d-1}
\frac{\chi_0}{\chi}\mu(1+\alpha_0)\left(\frac{\beta}{1+\beta}\right)^{1/2}A_c,
\end{equation}
\begin{equation}
\label{c8}
\nu_{d}^*=\frac{1}
{\sqrt{2}d(d+2)}\left(\frac{\overline{\sigma}}{\sigma}\right)^{d-1}
\frac{\chi_0}{\chi}\mu(1+\alpha_0)\left(\frac{\beta}{1+\beta}\right)^{3/2}
\left[A_d-(d+2)\frac{1+\beta}{\beta} A_c\right],
\end{equation}
where
\begin{eqnarray}
\label{c9}
A_c&=& (d+2)(1+2\lambda)+\mu(1+\beta)\Big\{(d+2)(1-\alpha_0)
-[(11+d)\alpha_0-5d-7]\lambda\beta^{-1}\Big\}+3(d+3)\lambda^2\beta^{-1}\nonumber\\
& &+2\mu^2\left(2\alpha_0^{2}-\frac{d+3}{2}\alpha
_{12}+d+1\right)\beta^{-1}(1+\beta)^2- (d+2)\beta^{-1}(1+\beta),
\end{eqnarray}
\begin{eqnarray}
\label{c10}
A_d&=&2\mu^2\left(\frac{1+\beta}{\beta}\right)^{2}
\left(2\alpha_0^{2}-\frac{d+3}{2}\alpha_0+d+1\right)
\big[d+5+(d+2)\beta\big]-\mu(1+\beta) \Big\{\lambda\beta^{-2}[(d+5)+(d+2)\beta]
\nonumber\\
& & \times
[(11+d)\alpha_0
-5d-7]-\beta^{-1}[20+d(15-7\alpha_0)+d^2(1-\alpha_0)-28\alpha_0] -(d+2)^2(1-\alpha_0)\Big\}
\nonumber\\
& & +3(d+3)\lambda^2\beta^{-2}[d+5+(d+2)\beta]+ 2\lambda\beta^{-1}[24+11d+d^2+(d+2)^2\beta]
\nonumber\\
& & +(d+2)\beta^{-1} [d+3+(d+8)\beta]-(d+2)(1+\beta)\beta^{-2}
[d+3+(d+2)\beta],\nonumber\\
\end{eqnarray}
Here, $\lambda=\mu_0\left(1-T/T_0\right)$.
\end{widetext}



\begin{thebibliography}{40}%
	\makeatletter
	\providecommand \@ifxundefined [1]{%
		\@ifx{#1\undefined}
	}%
	\providecommand \@ifnum [1]{%
		\ifnum #1\expandafter \@firstoftwo
		\else \expandafter \@secondoftwo
		\fi
	}%
	\providecommand \@ifx [1]{%
		\ifx #1\expandafter \@firstoftwo
		\else \expandafter \@secondoftwo
		\fi
	}%
	\providecommand \natexlab [1]{#1}%
	\providecommand \enquote  [1]{``#1''}%
	\providecommand \bibnamefont  [1]{#1}%
	\providecommand \bibfnamefont [1]{#1}%
	\providecommand \citenamefont [1]{#1}%
	\providecommand \href@noop [0]{\@secondoftwo}%
	\providecommand \href [0]{\begingroup \@sanitize@url \@href}%
	\providecommand \@href[1]{\@@startlink{#1}\@@href}%
	\providecommand \@@href[1]{\endgroup#1\@@endlink}%
	\providecommand \@sanitize@url [0]{\catcode `\\12\catcode `\$12\catcode
		`\&12\catcode `\#12\catcode `\^12\catcode `\_12\catcode `\%12\relax}%
	\providecommand \@@startlink[1]{}%
	\providecommand \@@endlink[0]{}%
	\providecommand \url  [0]{\begingroup\@sanitize@url \@url }%
	\providecommand \@url [1]{\endgroup\@href {#1}{\urlprefix }}%
	\providecommand \urlprefix  [0]{URL }%
	\providecommand \Eprint [0]{\href }%
	\providecommand \doibase [0]{https://doi.org/}%
	\providecommand \selectlanguage [0]{\@gobble}%
	\providecommand \bibinfo  [0]{\@secondoftwo}%
	\providecommand \bibfield  [0]{\@secondoftwo}%
	\providecommand \translation [1]{[#1]}%
	\providecommand \BibitemOpen [0]{}%
	\providecommand \bibitemStop [0]{}%
	\providecommand \bibitemNoStop [0]{.\EOS\space}%
	\providecommand \EOS [0]{\spacefactor3000\relax}%
	\providecommand \BibitemShut  [1]{\csname bibitem#1\endcsname}%
	\let\auto@bib@innerbib\@empty
	\bibitem [{\citenamefont
		{Smoluchowski}(1906{\natexlab{a}})}]{Smoluchowski1906a}%
	\BibitemOpen
	\bibfield  {author} {\bibinfo {author} {\bibfnamefont {M.}~\bibnamefont
			{Smoluchowski}},\ }\bibfield  {title} {\bibinfo {title} {Sur le chemin moyen
			parcouru par les molecules d'un gaz et sur son rapport avec la th\'eorie de
			la diffusion},\ }\href {https://archive.org/details/bulletininternat1906pols}
	{\bibfield  {journal} {\bibinfo  {journal} {Bulletin International d'Academie
				des Sciences de Cracovie, Classe des Sciences Math\'ematiques et Naturelles}\
			,\ \bibinfo {pages} {202}} (\bibinfo {year}
		{1906}{\natexlab{a}})}\BibitemShut {NoStop}%
	\bibitem [{\citenamefont
		{Smoluchowski}(1906{\natexlab{b}})}]{Smoluchowski1906b}%
	\BibitemOpen
	\bibfield  {author} {\bibinfo {author} {\bibfnamefont {M.}~\bibnamefont
			{Smoluchowski}},\ }\bibfield  {title} {\bibinfo {title} {Zur kinetischen
			theorie der brownschen molekularbewegung und der suspensionen},\ }\href
	{https://doi.org/10.1002/andp.19063261405} {\bibfield  {journal} {\bibinfo
			{journal} {Annalen der Physik}\ }\textbf {\bibinfo {volume} {326}},\ \bibinfo
		{pages} {756} (\bibinfo {year} {1906}{\natexlab{b}})}\BibitemShut {NoStop}%
	\bibitem [{\citenamefont {Einstein}(1905)}]{Einstein1905}%
	\BibitemOpen
	\bibfield  {author} {\bibinfo {author} {\bibfnamefont {A.}~\bibnamefont
			{Einstein}},\ }\bibfield  {title} {\bibinfo {title} {\"{U}ber einen die
			erzeugung und verwandlung des lichtes betreffenden heuristischen
			gesichtspunkt},\ }\href {https://doi.org/10.1002/andp.19053220607} {\bibfield
		{journal} {\bibinfo  {journal} {Annalen der Physik}\ }\textbf {\bibinfo
			{volume} {322}},\ \bibinfo {pages} {132} (\bibinfo {year}
		{1905})}\BibitemShut {NoStop}%
	\bibitem [{\citenamefont {Reif}(1965)}]{Reif1965}%
	\BibitemOpen
	\bibfield  {author} {\bibinfo {author} {\bibfnamefont {F.}~\bibnamefont
			{Reif}},\ }\href@noop {} {\emph {\bibinfo {title} {Fundamentals of
				Statistical and Thermal Physics}}}\ (\bibinfo  {publisher} {McGraw-Hill},\
	\bibinfo {year} {1965})\BibitemShut {NoStop}%
	\bibitem [{\citenamefont {Jeans}(1904)}]{Jeans1904}%
	\BibitemOpen
	\bibfield  {author} {\bibinfo {author} {\bibfnamefont {J.~H.}\ \bibnamefont
			{Jeans}},\ }\bibfield  {title} {\bibinfo {title} {{LXX}. {T}he persistence of
			molecular velocities in the kinetic theory of gases},\ }\href
	{https://doi.org/10.1080/14786440409463242} {\bibfield  {journal} {\bibinfo
			{journal} {The London, Edinburgh, and Dublin Philosophical Magazine and
				Journal of Science}\ }\textbf {\bibinfo {volume} {8}},\ \bibinfo {pages}
		{700} (\bibinfo {year} {1904})}\BibitemShut {NoStop}%
	\bibitem [{\citenamefont {Jeans}(2009)}]{Jeans1940}%
	\BibitemOpen
	\bibfield  {author} {\bibinfo {author} {\bibfnamefont {J.~H.}\ \bibnamefont
			{Jeans}},\ }\href@noop {} {\emph {\bibinfo {title} {An Introduction to the
				Kinetic Theory of Gases}}}\ (\bibinfo  {publisher} {Cambridge University
		Press, Cambridge, UK},\ \bibinfo {year} {2009})\BibitemShut {NoStop}%
	\bibitem [{\citenamefont {Yang}(1949)}]{Yang1949}%
	\BibitemOpen
	\bibfield  {author} {\bibinfo {author} {\bibfnamefont {L.~M.}\ \bibnamefont
			{Yang}},\ }\bibfield  {title} {\bibinfo {title} {Kinetic theory of diffusion
			in gases and liquids {I}. {D}iffusion and the {B}rownian motion},\ }\href
	{https://doi.org/10.1098/rspa.1949.0089} {\bibfield  {journal} {\bibinfo
			{journal} {Proc. R. Soc. Lond. A}\ }\textbf {\bibinfo {volume} {198}},\
		\bibinfo {pages} {94} (\bibinfo {year} {1949})}\BibitemShut {NoStop}%
	\bibitem [{\citenamefont {Yuste}\ \emph {et~al.}(2024)\citenamefont {Yuste},
		\citenamefont {G\'omez~Gonz\'alez},\ and\ \citenamefont
		{Garz\'o}}]{Yuste2024}%
	\BibitemOpen
	\bibfield  {author} {\bibinfo {author} {\bibfnamefont {S.~B.}\ \bibnamefont
			{Yuste}}, \bibinfo {author} {\bibfnamefont {R.}~\bibnamefont
			{G\'omez~Gonz\'alez}},\ and\ \bibinfo {author} {\bibfnamefont
			{V.}~\bibnamefont {Garz\'o}},\ }\bibfield  {title} {\bibinfo {title} {Gaseous
			diffusion as a correlated random walk},\ }\href
	{https://doi.org/10.1103/PhysRevE.110.014102} {\bibfield  {journal} {\bibinfo
			{journal} {Phys. Rev. E}\ }\textbf {\bibinfo {volume} {110}},\ \bibinfo
		{pages} {014102} (\bibinfo {year} {2024})}\BibitemShut {NoStop}%
	\bibitem [{\citenamefont {Chapman}\ \emph {et~al.}(1970)\citenamefont
		{Chapman}, \citenamefont {Cowling},\ and\ \citenamefont
		{Cercignani}}]{CCC70}%
	\BibitemOpen
	\bibfield  {author} {\bibinfo {author} {\bibfnamefont {S.}~\bibnamefont
			{Chapman}}, \bibinfo {author} {\bibfnamefont {T.}~\bibnamefont {Cowling}},\
		and\ \bibinfo {author} {\bibfnamefont {C.}~\bibnamefont {Cercignani}},\
	}\href {https://books.google.es/books?id=ZloRtgEACAAJ} {\emph {\bibinfo
			{title} {The Mathematical Theory of Non-uniform Gases}}}\ (\bibinfo
	{publisher} {Cambridge University Press, Cambridge, UK},\ \bibinfo {year}
	{1970})\BibitemShut {NoStop}%
	\bibitem [{\citenamefont {Brilliantov}\ and\ \citenamefont
		{P\"oschel}(2004)}]{BP04}%
	\BibitemOpen
	\bibfield  {author} {\bibinfo {author} {\bibfnamefont {N.~V.}\ \bibnamefont
			{Brilliantov}}\ and\ \bibinfo {author} {\bibfnamefont {T.}~\bibnamefont
			{P\"oschel}},\ }\href
	{https://doi.org/10.1093/acprof:oso/9780198530381.001.0001} {\emph {\bibinfo
			{title} {Kinetic Theory of Granular Gases}}}\ (\bibinfo  {publisher} {Oxford
		University Press},\ \bibinfo {year} {2004})\BibitemShut {NoStop}%
	\bibitem [{\citenamefont {Garz\'o}(2019)}]{GarzoBook19}%
	\BibitemOpen
	\bibfield  {author} {\bibinfo {author} {\bibfnamefont {V.}~\bibnamefont
			{Garz\'o}},\ }\href@noop {} {\emph {\bibinfo {title} {Granular Gaseous
				Flows}}}\ (\bibinfo  {publisher} {Springer Nature, Cham},\ \bibinfo {year}
	{2019})\BibitemShut {NoStop}%
	\bibitem [{\citenamefont {Goldhirsch}\ and\ \citenamefont
		{Zanetti}(1993)}]{GZ93}%
	\BibitemOpen
	\bibfield  {author} {\bibinfo {author} {\bibfnamefont {I.}~\bibnamefont
			{Goldhirsch}}\ and\ \bibinfo {author} {\bibfnamefont {G.}~\bibnamefont
			{Zanetti}},\ }\bibfield  {title} {\bibinfo {title} {Clustering instability in
			dissipative gases},\ }\href@noop {} {\bibfield  {journal} {\bibinfo
			{journal} {Phys. Rev. Lett.}\ }\textbf {\bibinfo {volume} {70}},\ \bibinfo
		{pages} {1619} (\bibinfo {year} {1993})}\BibitemShut {NoStop}%
	\bibitem [{\citenamefont {McNamara}(1993)}]{M93}%
	\BibitemOpen
	\bibfield  {author} {\bibinfo {author} {\bibfnamefont {S.}~\bibnamefont
			{McNamara}},\ }\bibfield  {title} {\bibinfo {title} {Hydrodynamic modes of a
			uniform granular medium},\ }\href@noop {} {\bibfield  {journal} {\bibinfo
			{journal} {Phys. Fluids A}\ }\textbf {\bibinfo {volume} {5}},\ \bibinfo
		{pages} {3056} (\bibinfo {year} {1993})}\BibitemShut {NoStop}%
	\bibitem [{\citenamefont {Brey}\ \emph {et~al.}(1998)\citenamefont {Brey},
		\citenamefont {Dufty}, \citenamefont {Kim},\ and\ \citenamefont
		{Santos}}]{BDKS98}%
	\BibitemOpen
	\bibfield  {author} {\bibinfo {author} {\bibfnamefont {J.~J.}\ \bibnamefont
			{Brey}}, \bibinfo {author} {\bibfnamefont {J.~W.}\ \bibnamefont {Dufty}},
		\bibinfo {author} {\bibfnamefont {C.~S.}\ \bibnamefont {Kim}},\ and\ \bibinfo
		{author} {\bibfnamefont {A.}~\bibnamefont {Santos}},\ }\bibfield  {title}
	{\bibinfo {title} {Hydrodynamics for granular flows at low density},\
	}\href@noop {} {\bibfield  {journal} {\bibinfo  {journal} {Phys. Rev. E}\
		}\textbf {\bibinfo {volume} {58}},\ \bibinfo {pages} {4638} (\bibinfo {year}
		{1998})}\BibitemShut {NoStop}%
	\bibitem [{\citenamefont {Brey}\ \emph {et~al.}(1999)\citenamefont {Brey},
		\citenamefont {Ruiz-Montero},\ and\ \citenamefont {Cubero}}]{BRC99}%
	\BibitemOpen
	\bibfield  {author} {\bibinfo {author} {\bibfnamefont {J.~J.}\ \bibnamefont
			{Brey}}, \bibinfo {author} {\bibfnamefont {M.~J.}\ \bibnamefont
			{Ruiz-Montero}},\ and\ \bibinfo {author} {\bibfnamefont {D.}~\bibnamefont
			{Cubero}},\ }\bibfield  {title} {\bibinfo {title} {Origin of density
			clustering in a freely evolving granular gas},\ }\href@noop {} {\bibfield
		{journal} {\bibinfo  {journal} {Phys. Rev. E}\ }\textbf {\bibinfo {volume}
			{60}},\ \bibinfo {pages} {3150} (\bibinfo {year} {1999})}\BibitemShut
	{NoStop}%
	\bibitem [{\citenamefont {Garz\'o}(2005)}]{G05}%
	\BibitemOpen
	\bibfield  {author} {\bibinfo {author} {\bibfnamefont {V.}~\bibnamefont
			{Garz\'o}},\ }\bibfield  {title} {\bibinfo {title} {Instabilities in a free
			granular fluid described by the {E}nskog equation},\ }\href@noop {}
	{\bibfield  {journal} {\bibinfo  {journal} {Phys. Rev. E}\ }\textbf {\bibinfo
			{volume} {72}},\ \bibinfo {pages} {021106} (\bibinfo {year}
		{2005})}\BibitemShut {NoStop}%
	\bibitem [{\citenamefont {Garz\'o}\ \emph {et~al.}(2006)\citenamefont
		{Garz\'o}, \citenamefont {Montanero},\ and\ \citenamefont {Dufty}}]{GMD06}%
	\BibitemOpen
	\bibfield  {author} {\bibinfo {author} {\bibfnamefont {V.}~\bibnamefont
			{Garz\'o}}, \bibinfo {author} {\bibfnamefont {J.~M.}\ \bibnamefont
			{Montanero}},\ and\ \bibinfo {author} {\bibfnamefont {J.~W.}\ \bibnamefont
			{Dufty}},\ }\bibfield  {title} {\bibinfo {title} {Mass and heat fluxes for a
			binary granular mixture at low density},\ }\href@noop {} {\bibfield
		{journal} {\bibinfo  {journal} {Phys. Fluids}\ }\textbf {\bibinfo {volume}
			{18}},\ \bibinfo {pages} {083305} (\bibinfo {year} {2006})}\BibitemShut
	{NoStop}%
	\bibitem [{\citenamefont {Mitrano}\ \emph {et~al.}(2012)\citenamefont
		{Mitrano}, \citenamefont {Garz\'o}, \citenamefont {Hilger}, \citenamefont
		{Ewasko},\ and\ \citenamefont {Hrenya}}]{MGHEH12}%
	\BibitemOpen
	\bibfield  {author} {\bibinfo {author} {\bibfnamefont {P.~P.}\ \bibnamefont
			{Mitrano}}, \bibinfo {author} {\bibfnamefont {V.}~\bibnamefont {Garz\'o}},
		\bibinfo {author} {\bibfnamefont {A.~M.}\ \bibnamefont {Hilger}}, \bibinfo
		{author} {\bibfnamefont {C.~J.}\ \bibnamefont {Ewasko}},\ and\ \bibinfo
		{author} {\bibfnamefont {C.~M.}\ \bibnamefont {Hrenya}},\ }\bibfield  {title}
	{\bibinfo {title} {Assessing a hydrodynamic description for instabilities in
			highly dissipative, freely cooling granular gases},\ }\href@noop {}
	{\bibfield  {journal} {\bibinfo  {journal} {Phys. Rev. E}\ }\textbf {\bibinfo
			{volume} {85}},\ \bibinfo {pages} {041303} (\bibinfo {year}
		{2012})}\BibitemShut {NoStop}%
	\bibitem [{\citenamefont {Brey}\ and\ \citenamefont
		{Ruiz-Montero}(2013)}]{BR13}%
	\BibitemOpen
	\bibfield  {author} {\bibinfo {author} {\bibfnamefont {J.~J.}\ \bibnamefont
			{Brey}}\ and\ \bibinfo {author} {\bibfnamefont {M.~J.}\ \bibnamefont
			{Ruiz-Montero}},\ }\bibfield  {title} {\bibinfo {title} {Shearing instability
			of a dilute granular mixture},\ }\href@noop {} {\bibfield  {journal}
		{\bibinfo  {journal} {Phys. Rev. E}\ }\textbf {\bibinfo {volume} {87}},\
		\bibinfo {pages} {022210} (\bibinfo {year} {2013})}\BibitemShut {NoStop}%
	\bibitem [{\citenamefont {Mitrano}\ \emph {et~al.}(2014)\citenamefont
		{Mitrano}, \citenamefont {Garz\'o},\ and\ \citenamefont {Hrenya}}]{MGH14}%
	\BibitemOpen
	\bibfield  {author} {\bibinfo {author} {\bibfnamefont {P.~P.}\ \bibnamefont
			{Mitrano}}, \bibinfo {author} {\bibfnamefont {V.}~\bibnamefont {Garz\'o}},\
		and\ \bibinfo {author} {\bibfnamefont {C.~M.}\ \bibnamefont {Hrenya}},\
	}\bibfield  {title} {\bibinfo {title} {Instabilities in granular binary
			mixtures at moderate density},\ }\href@noop {} {\bibfield  {journal}
		{\bibinfo  {journal} {Phys. Rev. E}\ }\textbf {\bibinfo {volume} {89}},\
		\bibinfo {pages} {020201 (R)} (\bibinfo {year} {2014})}\BibitemShut {NoStop}%
	\bibitem [{\citenamefont {Garz\'o}(2015)}]{G15}%
	\BibitemOpen
	\bibfield  {author} {\bibinfo {author} {\bibfnamefont {V.}~\bibnamefont
			{Garz\'o}},\ }\bibfield  {title} {\bibinfo {title} {Stability of freely
			cooling granular mixtures at moderate densities},\ }\href@noop {} {\bibfield
		{journal} {\bibinfo  {journal} {Chaos, Solitons and Fractals}\ }\textbf
		{\bibinfo {volume} {81}},\ \bibinfo {pages} {497} (\bibinfo {year}
		{2015})}\BibitemShut {NoStop}%
	\bibitem [{\citenamefont {Garz\'o}\ and\ \citenamefont
		{Montanero}(2004)}]{GM04}%
	\BibitemOpen
	\bibfield  {author} {\bibinfo {author} {\bibfnamefont {V.}~\bibnamefont
			{Garz\'o}}\ and\ \bibinfo {author} {\bibfnamefont {J.~M.}\ \bibnamefont
			{Montanero}},\ }\bibfield  {title} {\bibinfo {title} {Diffusion of impurities
			in a granular gas},\ }\href@noop {} {\bibfield  {journal} {\bibinfo
			{journal} {Phys. Rev. E}\ }\textbf {\bibinfo {volume} {69}},\ \bibinfo
		{pages} {021301} (\bibinfo {year} {2004})}\BibitemShut {NoStop}%
	\bibitem [{\citenamefont {Garz\'o}\ and\ \citenamefont
		{Vega~Reyes}(2009)}]{GV09}%
	\BibitemOpen
	\bibfield  {author} {\bibinfo {author} {\bibfnamefont {V.}~\bibnamefont
			{Garz\'o}}\ and\ \bibinfo {author} {\bibfnamefont {F.}~\bibnamefont
			{Vega~Reyes}},\ }\bibfield  {title} {\bibinfo {title} {Mass transport of
			impurities in a moderately dense granular gas},\ }\href@noop {} {\bibfield
		{journal} {\bibinfo  {journal} {Phys. Rev. E}\ }\textbf {\bibinfo {volume}
			{79}},\ \bibinfo {pages} {041303} (\bibinfo {year} {2009})}\BibitemShut
	{NoStop}%
	\bibitem [{\citenamefont {Bird}(1994)}]{Bird94}%
	\BibitemOpen
	\bibfield  {author} {\bibinfo {author} {\bibfnamefont {G.~A.}\ \bibnamefont
			{Bird}},\ }\href@noop {} {\emph {\bibinfo {title} {Molecular Gas Dynamics and
				the Direct Simulation Monte Carlo of Gas Flows}}}\ (\bibinfo  {publisher}
	{Clarendon Press, Oxford},\ \bibinfo {year} {1994})\BibitemShut {NoStop}%
	\bibitem [{\citenamefont {Abad}\ \emph {et~al.}(2022)\citenamefont {Abad},
		\citenamefont {Yuste},\ and\ \citenamefont {Garz\'o}}]{Abad2022}%
	\BibitemOpen
	\bibfield  {author} {\bibinfo {author} {\bibfnamefont {E.}~\bibnamefont
			{Abad}}, \bibinfo {author} {\bibfnamefont {S.~B.}\ \bibnamefont {Yuste}},\
		and\ \bibinfo {author} {\bibfnamefont {V.}~\bibnamefont {Garz\'o}},\
	}\bibfield  {title} {\bibinfo {title} {On the mean square displacement of
			intruders in freely cooling granular gases},\ }\href
	{https://doi.org/10.1007/s10035-022-01256-0} {\bibfield  {journal} {\bibinfo
			{journal} {Granular Matter}\ }\textbf {\bibinfo {volume} {24}},\ \bibinfo
		{pages} {111} (\bibinfo {year} {2022})}\BibitemShut {NoStop}%
	\bibitem [{\citenamefont {Haff}(1983)}]{H83}%
	\BibitemOpen
	\bibfield  {author} {\bibinfo {author} {\bibfnamefont {P.~K.}\ \bibnamefont
			{Haff}},\ }\bibfield  {title} {\bibinfo {title} {Grain flow as a
			fluid-mechanical phenomenon},\ }\href@noop {} {\bibfield  {journal} {\bibinfo
			{journal} {J. Fluid Mech.}\ }\textbf {\bibinfo {volume} {134}},\ \bibinfo
		{pages} {401} (\bibinfo {year} {1983})}\BibitemShut {NoStop}%
	\bibitem [{\citenamefont {Garz\'o}\ and\ \citenamefont {Dufty}(1999)}]{GD99b}%
	\BibitemOpen
	\bibfield  {author} {\bibinfo {author} {\bibfnamefont {V.}~\bibnamefont
			{Garz\'o}}\ and\ \bibinfo {author} {\bibfnamefont {J.~W.}\ \bibnamefont
			{Dufty}},\ }\bibfield  {title} {\bibinfo {title} {Homogeneous cooling state
			for a granular mixture},\ }\href@noop {} {\bibfield  {journal} {\bibinfo
			{journal} {Phys. Rev. E}\ }\textbf {\bibinfo {volume} {60}},\ \bibinfo
		{pages} {5706} (\bibinfo {year} {1999})}\BibitemShut {NoStop}%
	\bibitem [{\citenamefont {Paik}(2014)}]{Paik2014}%
	\BibitemOpen
	\bibfield  {author} {\bibinfo {author} {\bibfnamefont {S.~T.}\ \bibnamefont
			{Paik}},\ }\bibfield  {title} {\bibinfo {title} {{Is the mean free path the
				mean of a distribution?}},\ }\href {https://doi.org/10.1119/1.4869185}
	{\bibfield  {journal} {\bibinfo  {journal} {American Journal of Physics}\
		}\textbf {\bibinfo {volume} {82}},\ \bibinfo {pages} {602} (\bibinfo {year}
		{2014})}\BibitemShut {NoStop}%
	\bibitem [{\citenamefont {Rubinstein}\ and\ \citenamefont
		{Colby}(2003)}]{Rubinstein2003}%
	\BibitemOpen
	\bibfield  {author} {\bibinfo {author} {\bibfnamefont {M.}~\bibnamefont
			{Rubinstein}}\ and\ \bibinfo {author} {\bibfnamefont {R.~H.}\ \bibnamefont
			{Colby}},\ }\href {https://doi.org/10.1093/oso/9780198520597.001.0001} {\emph
		{\bibinfo {title} {Polymer Physics}}}\ (\bibinfo  {publisher} {Oxford
		University Press},\ \bibinfo {year} {2003})\BibitemShut {NoStop}%
	\bibitem [{\citenamefont {Grosberg}\ and\ \citenamefont
		{Khokhlov}(1994)}]{Grosberg1994}%
	\BibitemOpen
	\bibfield  {author} {\bibinfo {author} {\bibfnamefont {A.~Y.}\ \bibnamefont
			{Grosberg}}\ and\ \bibinfo {author} {\bibfnamefont {A.~R.}\ \bibnamefont
			{Khokhlov}},\ }\href@noop {} {\emph {\bibinfo {title} {Statistical Physics of
				Macromolecules}}}\ (\bibinfo  {publisher} {American Institute of Physics, New
		York},\ \bibinfo {year} {1994})\BibitemShut {NoStop}%
	\bibitem [{\citenamefont {Bodrova}\ \emph {et~al.}(2015)\citenamefont
		{Bodrova}, \citenamefont {Chechkin}, \citenamefont {Cherstvy},\ and\
		\citenamefont {Metzler}}]{Bodrova2015}%
	\BibitemOpen
	\bibfield  {author} {\bibinfo {author} {\bibfnamefont {A.}~\bibnamefont
			{Bodrova}}, \bibinfo {author} {\bibfnamefont {A.~V.}\ \bibnamefont
			{Chechkin}}, \bibinfo {author} {\bibfnamefont {A.~G.}\ \bibnamefont
			{Cherstvy}},\ and\ \bibinfo {author} {\bibfnamefont {R.}~\bibnamefont
			{Metzler}},\ }\bibfield  {title} {\bibinfo {title} {Quantifying non-ergodic
			dynamics of force-free granular gases},\ }\href
	{https://doi.org/10.1039/C5CP02824H} {\bibfield  {journal} {\bibinfo
			{journal} {Phys. Chem. Chem. Phys.}\ }\textbf {\bibinfo {volume} {17}},\
		\bibinfo {pages} {21791} (\bibinfo {year} {2015})}\BibitemShut {NoStop}%
	\bibitem [{\citenamefont {Bodrova}(2024)}]{Bodrova2024}%
	\BibitemOpen
	\bibfield  {author} {\bibinfo {author} {\bibfnamefont {A.~S.}\ \bibnamefont
			{Bodrova}},\ }\bibfield  {title} {\bibinfo {title} {Diffusion in
			multicomponent granular mixtures},\ }\href
	{https://doi.org/10.1103/PhysRevE.109.024903} {\bibfield  {journal} {\bibinfo
			{journal} {Phys. Rev. E}\ }\textbf {\bibinfo {volume} {109}},\ \bibinfo
		{pages} {024903} (\bibinfo {year} {2024})}\BibitemShut {NoStop}%
	\bibitem [{\citenamefont {Bodrova}\ and\ \citenamefont
		{Osinsky}(2025)}]{Bodrova2025}%
	\BibitemOpen
	\bibfield  {author} {\bibinfo {author} {\bibfnamefont {A.~S.}\ \bibnamefont
			{Bodrova}}\ and\ \bibinfo {author} {\bibfnamefont {A.~I.}\ \bibnamefont
			{Osinsky}},\ }\bibfield  {title} {\bibinfo {title} {Anomalous diffusion in
			polydisperse granular gases},\ }\href
	{https://doi.org/10.1103/PhysRevE.111.035402} {\bibfield  {journal} {\bibinfo
			{journal} {Phys. Rev. E}\ }\textbf {\bibinfo {volume} {111}},\ \bibinfo
		{pages} {035402} (\bibinfo {year} {2025})}\BibitemShut {NoStop}%
	\bibitem [{\citenamefont {Bodrova}\ \emph {et~al.}(2012)\citenamefont
		{Bodrova}, \citenamefont {Dubey}, \citenamefont {Puri},\ and\ \citenamefont
		{Brilliantov}}]{Bodrova2012}%
	\BibitemOpen
	\bibfield  {author} {\bibinfo {author} {\bibfnamefont {A.}~\bibnamefont
			{Bodrova}}, \bibinfo {author} {\bibfnamefont {A.~K.}\ \bibnamefont {Dubey}},
		\bibinfo {author} {\bibfnamefont {S.}~\bibnamefont {Puri}},\ and\ \bibinfo
		{author} {\bibfnamefont {N.}~\bibnamefont {Brilliantov}},\ }\bibfield
	{title} {\bibinfo {title} {Intermediate regimes in granular brownian motion:
			Superdiffusion and subdiffusion},\ }\href
	{https://doi.org/10.1103/PhysRevLett.109.178001} {\bibfield  {journal}
		{\bibinfo  {journal} {Phys. Rev. Lett.}\ }\textbf {\bibinfo {volume} {109}},\
		\bibinfo {pages} {178001} (\bibinfo {year} {2012})}\BibitemShut {NoStop}%
	\bibitem [{\citenamefont {Gómez~González}\ \emph {et~al.}(2024)\citenamefont
		{Gómez~González}, \citenamefont {Garzó}, \citenamefont {Brito},\ and\
		\citenamefont {Soto}}]{GGGBS24}%
	\BibitemOpen
	\bibfield  {author} {\bibinfo {author} {\bibfnamefont {R.}~\bibnamefont
			{Gómez~González}}, \bibinfo {author} {\bibfnamefont {V.}~\bibnamefont
			{Garzó}}, \bibinfo {author} {\bibfnamefont {R.}~\bibnamefont {Brito}},\ and\
		\bibinfo {author} {\bibfnamefont {R.}~\bibnamefont {Soto}},\ }\bibfield
	{title} {\bibinfo {title} {Diffusion of impurities in a moderately dense
			confined granular gas},\ }\href {https://doi.org/10.1063/5.0245373}
	{\bibfield  {journal} {\bibinfo  {journal} {Physics of Fluids}\ }\textbf
		{\bibinfo {volume} {36}},\ \bibinfo {pages} {123387} (\bibinfo {year}
		{2024})}\BibitemShut {NoStop}%
	\bibitem [{\citenamefont {Santos}\ and\ \citenamefont {Blinde}()}]{Santos2008}%
	\BibitemOpen
	\bibfield  {author} {\bibinfo {author} {\bibfnamefont {A.}~\bibnamefont
			{Santos}}\ and\ \bibinfo {author} {\bibfnamefont {S.~M.}\ \bibnamefont
			{Blinde}},\ }\href
	{https://demonstrations.wolfram.com/InelasticCollisionsOfTwoSpheres/}
	{\bibinfo {title} {Inelastic collisions of two spheres. {W}olfram
			demonstrations project.}}\BibitemShut {Stop}%
	\bibitem [{\citenamefont {Santos}(2016)}]{Santos2016book}%
	\BibitemOpen
	\bibfield  {author} {\bibinfo {author} {\bibfnamefont {A.}~\bibnamefont
			{Santos}},\ }\href {https://books.google.es/books?id=fKoqDAAAQBAJ} {\emph
		{\bibinfo {title} {A Concise Course on the Theory of Classical Liquids:
				Basics and Selected Topics}}},\ Lecture Notes in Physics\ (\bibinfo
	{publisher} {Springer International Publishing},\ \bibinfo {year}
	{2016})\BibitemShut {NoStop}%
	\bibitem [{\citenamefont {Garz\'o}\ and\ \citenamefont {Santos}(2003)}]{GS03}%
	\BibitemOpen
	\bibfield  {author} {\bibinfo {author} {\bibfnamefont {V.}~\bibnamefont
			{Garz\'o}}\ and\ \bibinfo {author} {\bibfnamefont {A.}~\bibnamefont
			{Santos}},\ }\href@noop {} {\emph {\bibinfo {title} {Kinetic Theory of Gases
				in Shear Flows. Nonlinear Transport}}}\ (\bibinfo  {publisher} {Springer,
		Netherlands},\ \bibinfo {year} {2003})\BibitemShut {NoStop}%
	\bibitem [{\citenamefont {Yuste}\ and\ \citenamefont
		{G\'omez~Gonz\'alez}()}]{datasetHCS}%
	\BibitemOpen
	\bibfield  {author} {\bibinfo {author} {\bibfnamefont {S.~B.}\ \bibnamefont
			{Yuste}}\ and\ \bibinfo {author} {\bibfnamefont {R.}~\bibnamefont
			{G\'omez~Gonz\'alez}},\ }\href {https://doi.org/10.5281/zenodo.15350988}
	{\bibinfo {title} {{Simulation data of tracer particles diffusing in a granular gas of inelastic hard spheres under homogeneous cooling state},{
				Zenodo (2025), doi:10.5281/zenodo.15350988}}}\BibitemShut {NoStop}%
	\bibitem [{\citenamefont {van Noije}\ and\ \citenamefont {Ernst}(1998)}]{NE98}%
	\BibitemOpen
	\bibfield  {author} {\bibinfo {author} {\bibfnamefont {T.~P.~C.}\
			\bibnamefont {van Noije}}\ and\ \bibinfo {author} {\bibfnamefont {M.~H.}\
			\bibnamefont {Ernst}},\ }\bibfield  {title} {\bibinfo {title} {Velocity
			distributions in homogeneous granular fluids: the free and heated case},\
	}\href@noop {} {\bibfield  {journal} {\bibinfo  {journal} {Granular Matter}\
		}\textbf {\bibinfo {volume} {1}},\ \bibinfo {pages} {57} (\bibinfo {year}
		{1998})}\BibitemShut {NoStop}%
\end{thebibliography}

%

\end{document}